\documentclass[12pt,a4paper]{article}
\pdfoutput=1

\usepackage{graphicx,array}
\usepackage{hyperref}
\usepackage[normalem]{ulem} 
\usepackage{cite}
\usepackage{color}
\usepackage{appendix}
\usepackage{epsfig}
\usepackage{latexsym}
\usepackage{amsmath}
\usepackage{amssymb}
\usepackage{bm}
\usepackage{hyperref}
\usepackage{slashed}
\usepackage{subfig}
\usepackage{multirow}
\usepackage{adjustbox}
\usepackage{colortbl}
\usepackage{float}
\usepackage{placeins}

\numberwithin{equation}{section}

\definecolor{darkblue}{cmyk}{1,0.3,0,0.2}
\definecolor{violet}{cmyk}{0,1,0,0.2}
\hypersetup{colorlinks, bookmarksnumbered, citecolor=darkblue, linkcolor=darkblue, pdfstartview=FitH, urlcolor=darkblue, linktocpage}

\newcommand{\be}{\begin{equation}}
\newcommand{\ee}{\end{equation}}
\newcommand{\bea}{\begin{eqnarray}}
\newcommand{\eea}{\end{eqnarray}}

\newcommand{\TeV}{\textrm{ TeV}}

\newcommand{\SM}{\textrm{SM}}
\newcommand{\gsim}{\lower.7ex\hbox{$\;\stackrel{\textstyle>}{\sim}\;$}}
\newcommand{\lsim}{\lower.7ex\hbox{$\;\stackrel{\textstyle<}{\sim}\;$}}

\newcommand{\LL}{\mathcal{L}}

\newcommand{\OO}{\mathcal{O}}

\newcommand{\NN}{\mathcal{N}}

\newcommand{\Br}{\mathcal{B}}

\newcommand{\ba} {\begin{eqnarray}}
\newcommand{\ea} {\end{eqnarray}}

\newcommand{\cA} {\mathcal A}
\newcommand{\cB} {\mathcal B}

\renewcommand{\Re}{\mathop{\rm Re}}
\renewcommand{\Im}{\mathop{\rm Im}}

\begin{document}
 
 {\large
\flushright TUM-HEP-1344-21\\ }

 \hfill

\vspace{1.0cm}

\begin{center}
{\LARGE\bf From B-meson anomalies to Kaon physics\\[0.5cm] with scalar leptoquarks}
\\ %\vspace*{0.5cm}

\bigskip\vspace{1cm}{
{\large \mbox{David Marzocca$^{a}$},  \mbox{Sokratis Trifinopoulos$^{a}$}, \mbox{Elena Venturini$^{b}$} }
} \\[7mm]
{\em $(a)$ INFN, Sezione di Trieste, SISSA, Via Bonomea 265, 34136, Trieste, Italy}  \\ 
{\em $(b)$ Technische Universit{\"a}t M{\"u}nchen, Physik-Department, James-Franck-Stra\ss e 1, 85748 Garching, Germany}\\ 

\vspace*{0.5cm}%\today
   
\end{center}
\vspace*{1.5cm}

\centerline{\large\bf Abstract}
\medskip\noindent 
In this work we study possible connections between $B$-meson anomalies and Kaon physics observables in the context of combined solutions with the singlet and triplet scalar leptoquarks $S_1$ and $S_3$.
By assuming a flavor structure for the leptoquark couplings dictated by a minimally broken $U(2)^5$ flavor symmetry we can make a sharp connection between these two classes of observables. We find that the bound on $\mathcal{B}(K^+ \rightarrow \pi^+ \nu\nu)$ from NA62 puts already some tension in the model, while the present limits on $\mathcal{B}(K_L \rightarrow \mu^+ \mu^-)$ and $\mu \to e$ conversion in nuclei can be saturated.
Relaxing instead the flavor assumption we study what values for $\mathcal{B}(K^+ \rightarrow\pi^+ \nu\nu)$, as well as for $\mathcal{B}(K_L \rightarrow\pi^0 \nu\nu)$ and $\mathcal{B}(K_{L,S} \rightarrow\mu^+ \mu^-)$, are viable compatibly with all other phenomenological constraints.

\vspace{0.3cm}

\newpage
\tableofcontents

%\newpage

%%%%%%%%%%%%%%%%%%%%%%%%%%%%%%%%%%
\section{Introduction}
\label{sec:intro}

The observed deviations from Standard Model (SM) predictions in semileptonic $B$-meson decays persist as some of the most significant experimental hints for the presence of possible New Physics (NP) beyond the SM at the TeV scale.
One set of deviations regards Lepton Flavor Universality (LFU) ratios of charged-current semileptonic $B$ decays between the third and lighter lepton families, $R(D^{(*)}) = \Br(B \to D^{(*)} \tau \nu) / \Br(B \to D^{(*)} \ell \nu)$ 
\cite{Lees:2012xj,Lees:2013uzd,Aaij:2015yra,Huschle:2015rga,Sato:2016svk,Hirose:2016wfn,Hirose:2017dxl,Aaij:2017uff,Aaij:2017deq,Siddi:2018avt,Belle:2019rba}, with a combined significance of approximately $3\sigma$ \cite{Amhis:2016xyh}.
Another set of deviations from the SM are observed in LFU ratios between second and first lepton families in neutral-current $B$ decays, $R_{K^{(*)}} = \Br(B \to K^{(*)} \mu^+ \mu^-) / \Br(B \to K^{(*)} e^+ e^-)$ \cite{Aaij:2014ora,Aaij:2017vbb,Aaij:2019wad,Abdesselam:2019wac,Aaij:2021vac}, as well as in $B_s \to \mu^+ \mu^-$, in angular observables of the $B \to K^* \mu^+ \mu^-$ process as well as in branching ratios of other decay processes which involve the $b \to s \mu^+ \mu^-$ transition \cite{Aaij:2013qta,Aaij:2015oid,Aaij:2015esa,Aaij:2017vad,Aaboud:2018mst,Aaij:2020nrf}.
The global significance of these deviations, obtained with very conservative estimates of SM uncertainties, is $3.9\sigma$ \cite{Lancierini:2021sdf}. On the other hand, global fits show that along the preferred directions in effective field theory (EFT) space the pulls from the SM can be even up to $5-7\sigma$ \cite{Alok:2019ufo,Altmannshofer:2021qrr,Geng:2021nhg,Alguero:2021anc,Carvunis:2021jga,Alguero:2019ptt,Ciuchini:2019usw}.

When attempting to address at the same time both sets of anomalies, leptoquark (LQ) mediators are by far the preferred candidates. This is mainly due to the fact that, while semileptonic operators required for the $B$-anomalies can be induced at the tree-level, four-quark and four-lepton operators, that are strongly constrained by meson mixing or LFV, are induced only at one loop and thus automatically suppressed.

An interesting scenario for a combined explanation of the anomalies involves the two scalar LQs $S_1 = ({\bf \bar 3}, {\bf 1}, 1/3)$ and $S_3 = ({\bf \bar 3}, {\bf 3}, 1/3)$ \cite{Crivellin:2017zlb,Buttazzo:2017ixm,Marzocca:2018wcf,Heeck:2018ntp,Arnan:2019olv,Crivellin:2019dwb,Saad:2020ihm,Crivellin:2020ukd,Gherardi:2020qhc,DaRold:2020bib,Bordone:2020lnb}.
Interestingly enough, the $S_1$ couplings to right-handed fermions allow also an explanation of the observed deviation in the muon anomalous magnetic moment $(g-2)_\mu$ \cite{Bennett:2006fi,Abi:2021gix,Aoyama:2020ynm}.
As a possible ultraviolet (UV) completion for this model, the two scalars could arise, together with the Higgs boson, as pseudo-Nambu-Goldstone bosons from a new strongly coupled sector at the multi-TeV scale, which would also address the hierarchy problem of the electroweak scale in a composite Higgs framework \cite{Marzocca:2018wcf,DaRold:2020bib} (see also \cite{Gripaios:2009dq,Gripaios:2014tna,Alvarez:2018gxs} for related works).

The solution to $B$-anomalies involves NP couplings to second and third generations of quarks and leptons, while the couplings to first generation could in principle be very small, which is also required by strong experimental constraints \cite{Davidson:1993qk,Dorsner:2016wpm,Gherardi:2019zil,Mandal:2019gff,Crivellin:2021egp}.
Nevertheless, the natural expectation is that NP should couple to all generations, possibly with some flavor structure dictated by a dynamical mechanism or a symmetry.
In light of this, a question one can pose is: \emph{what are the expected effects in Kaon physics and electron observables for models that address the $B$-anomalies?}
This is the main question we aim at addressing with this paper, in the context of $S_1 + S_3$ solutions. Analyses in the same spirit were performed in Refs.~ \cite{Bordone:2017lsy,Borsato:2018tcz}, in an EFT context, and in Ref.~\cite{Fajfer:2018bfj} for single leptoquarks and in connection with $R(K^{(*)})$ only.
Specifically, the golden channels of rare Kaon decays, $K^+ \to \pi^+ \nu \nu$ and $K_L \to \pi^0 \nu \nu$, are now actively being investigated by the NA62 \cite{NA62:2021zjw} and KOTO \cite{Tanabashi:2018oca} experiments, respectively, and substantial improvements are expected in the future. Furthermore, an improvement by a few orders of magnitude in sensitivity is expected for $\mu \to e$ conversion experiments COMET and Mu2e \cite{Kuno:2013mha,Ankenbrandt:2006zu,Knoepfel:2013ouy,Bartoszek:2014mya} as well as in $\mu \to 3 e$ from the Mu3e experiment at PSI \cite{Mu3e:2020gyw}. 
Can these experiments expect to observe a signal, possible related to the $B$-anomalies?

Our starting point is the analysis of how this setup can address the $B$-anomalies (as well as the $(g-2)_\mu$) performed in Ref.~\cite{Gherardi:2020qhc}. In that work several scenarios were considered and, for each of them, a global analysis was performed including all the relevant observables computed at one-loop accuracy using the complete one-loop matching between the two mediators and the SMEFT obtained in Ref.~\cite{Gherardi:2020det}.
In particular, two scenarios able to address both sets of $B$-anomalies were found:
\begin{itemize}
    \item {\bf LH couplings.} If the $S_1$ couplings to right-handed fermions are zero, then both charged and neutral-current $B$-anomalies can be addressed, while the muon magnetic moment deviation cannot. It was also observed that the preferred values of the couplings to second generation was compatible with the structure hinted to by an approximate $U(2)^5$ flavor symmetry \cite{Gherardi:2020qhc} (see also \cite{Buttazzo:2017ixm,Marzocca:2018wcf}).
    \item {\bf All couplings.} If, instead, all couplings are allowed, then both the $B$-anomalies and the $(g-2)_\mu$ can be addressed, but the coupling structure is not compatible with the $U(2)^5$ flavor symmetry since a large coupling to $c_R \tau_R$ is required, with a small coupling to $t_R \tau_R$ instead \cite{Gherardi:2020qhc}.
\end{itemize}
We extend our previous work by considering also Kaon and $D$ decays, as well as all processes sensitive to the $\mu \to e$ lepton flavor violating (LFV) transition.

To find possible connections between $B$-anomalies and Kaon physics we take two different approaches for the two scenarios listed above.
For the first scenario we impose from the beginning a concrete assumption on the flavor structure, in particular the $U(2)^5$ flavor symmetry~\cite{Barbieri:2011ci,Barbieri:2012uh,Barbieri:2012tu} and perform the first complete study of $B$-anomalies and Kaon physics with $S_1$ and $S_3$ within this context. The main feature of this approximate symmetry is that strict relations between the LQ coupling to first and second generations are predicted, implying that Kaon decays become strictly connected with $B$ decays. \par
Our choice of the flavor symmetry is motivated by the observation that the approximate $U(2)_q \times U(2)_{\ell}$ flavor symmetry, that acts on the light-generations of SM quarks and leptons and is a subset of $U(2)^5$, appears to provide a consistent picture of all low-energy data. In fact, it was shown in the general EFT context~\cite{Greljo:2015mma,Barbieri:2015yvd,Buttazzo:2017ixm,Bordone:2017anc,Fuentes-Martin:2019mun}, as well as also in concrete UV realizations~\cite{Barbieri:2016las,Marzocca:2018wcf,Greljo:2018tuh,Bordone:2018nbg,Trifinopoulos:2018rna,DiLuzio:2018zxy,Trifinopoulos:2019lyo,Gherardi:2020qhc}, that not only it can reproduce the observed hierarchies in the SM Yukawa sector, but also successfully control the strength of NP couplings allowing for sufficiently large effects in processes involving third-generation fermions. We mention that the flavor symmetry does not need to be a fundamental property of the UV theory, but it could arise as an accidental low-energy symmetry. \par
For the second scenario, since no evident flavor structure emerges from the $B$-anomalies fit when also right-handed couplings are included, we let vary the couplings relevant for $K \to \pi \nu \nu$, while keeping the other couplings fixed to the best-fit values required by the $B$-anomalies and muon magnetic moment. In this way we can find the allowed values for $K \to \pi \nu \nu$ that are compatible with the $B$-anomalies, in this general setup.

The paper is structured as follows. In Section~\ref{sec:setup} we briefly introduce the model, the setup, and the statistical tool used for the analysis.
In Section~\ref{sec:LQandU2} we discuss the structure of the LQ couplings predicted by the minimally broken $U(2)^5$ flavor symmetry and study the results of the global fit in this framework.
Section~\ref{sec:LQgeneral} is instead devoted to the study of rare Kaon decays and electron LFV processes in the general case where the flavor assumption is lifted.
We conclude in Section~\ref{sec:conclusions}. Details on the observables not discussed in \cite{Gherardi:2020qhc} are collected in Appendix~\ref{app:obs}.

%%%%%%%%%%%%%%%%%%%%%%%%%%%%%%%%%%
\section{Setup}
\label{sec:setup}

We consider in this work the two scalar LQs $S_1 = ({\bf \bar 3}, {\bf 1}, 1/3)$ and $S_3 = ({\bf \bar 3}, {\bf 3}, 1/3)$, where the quantum numbers under the SM gauge group $SU(3)_c \times SU(2)_L \times U(1)_Y$ are indicated. The Lagrangian to be added to the SM one, assuming baryon and lepton number conservation, is
\be\begin{split}
	\LL_{\text{LQ}} =& |D_\mu S_1|^2 + |D_\mu S_3|^2 - M_{1}^2 |S_1|^2 - M_{3} ^2 |S_3|^2 - V_{\text{LQ}}(S_1, S_3, H)+ \\
		& + \left( (\lambda^{1L})_{i\alpha} \, \overline{q^c}_i  \,\epsilon\, \ell_\alpha
			+ (\lambda^{1R})_{i\alpha} \, \overline{u^c}_i   e_\alpha  \right) S_1 
			+ (\lambda^{3L})_{i\alpha} \, \overline{q^c}_i \,\epsilon\,  \sigma^I \ell_\alpha S_3^I + \text{h.c.} ~,
	\label{eq:S1S3Model}
\end{split}\ee
where $\epsilon=i\sigma _2$, $(\lambda^{1L})_{i\alpha}, (\lambda^{1R})_{i\alpha}, (\lambda^{3L})_{i\alpha} \in \mathbb{C}$, and $V_{\text{LQ}}$ includes LQ self-couplings and interactions with the Higgs boson, which are omitted since they are not relevant for the phenomenology studied here.
We denote SM quark and lepton fields by $q_i$, $u_i$, $d_i$, $\ell _\alpha$, and $e_\alpha$, while the Higgs doublet is $H$. We adopt latin letters ($i,\,j,\,k,\,\dots$) for quark flavor indices and greek letters ($\alpha,\,\beta,\,\gamma,\,\dots$) for lepton flavor indices.
We work in the down-quark and charged-lepton mass eigenstate basis, where
\be
	q_i = \left( \begin{array}{c} V^*_{ji} u^j_L \\ d^i_L  \end{array}  \right)\,, \qquad
	\ell_\alpha = \left( \begin{array}{c} \nu^\alpha_L \\ e^\alpha_L \end{array} \right)~,
\ee
and $V$ is the CKM matrix. We use the same conventions as Refs.~\cite{Gherardi:2020det,Gherardi:2020qhc}, to which we refer for further details.

Our goal is to study the phenomenology of the $S_1+S_3$ model extending the list of observables already accounted for in Ref.~\cite{Gherardi:2020qhc} to all the other relevant Kaon, $D$-meson, and electron LFV ones. We do this by employing the same multistep procedure: the LQ model is matched at one-loop level into the SMEFT \cite{Gherardi:2020det}, which is then matched into the Low Energy EFT (LEFT) \cite{Jenkins:2017jig,Dekens:2019ept}; within any EFT the renormalization group evolution (RGE) is taken into account, as well as the one-loop rational contributions within the LEFT, in terms of whose coefficients the observables and pseudo-observables are expressed (barring the case of observables that are measured at the electroweak scale).

In our global analysis for the two LQs we add the following observables to those observables already studied in Ref.~\cite{Gherardi:2020qhc}: rare and LFV Kaon decays, $\epsilon_K'/\epsilon_K$, rare $D$-meson decays, $b\to d\ell\ell$ decays, $\mu\to e$ conversion in nuclei, and the neutron EDM.
In Tables~\ref{tab:obs}, \ref{tab:obsK}, \ref{tab:obs2}, and \ref{tab:ZcouplLEP}, we show the complete list of observables that we analyze, together with their SM predictions and experimental bounds.
In App.~\ref{app:obs} we collect details on the low-energy observables that were not considered in Ref.~\cite{Gherardi:2020qhc}, together with others which turn out to set relevant constraints in the fit, such as limits from $Z$ couplings measurements and $K\to \pi \nu\nu$, that we repeat for sake of completeness.
For all the observables, the full set of one-loop corrections is considered in the numerical analysis.

We perform a $\chi^2$ fit, thus defining the likelihood as
\be
	- 2 \log \LL \equiv \chi^2(\lambda_x, M_x) = \sum_i \frac{ \left(\OO_i(\lambda_x, M_x) - \mu_i\right)^2}{\sigma_i^2}~,
	\label{eq:Global_chiSQ}
\ee
where $\OO_i(\lambda_x, M_x)$ is the expression of the observable as function of the model parameters, $\mu_i$ its central measured value, and $\sigma_i$ the associated standard deviation. In the analysis presented in this paper, 73 observables are taken into account, for which, within the SM, the $\chi^2$ is $\chi^2_{\rm SM}=104.0$.
In each scenario we first find the best-fit point by minimizing the $\chi^2$. We then perform a numerical scan over the parameter space using a Markov Chain Monte Carlo algorithm (Hastings-Metropolis), to select points that are within the 68 or 95\% CL from the best-fit point, with final samples of size $\mathcal{O}(10^4)$. These scans are used to obtain preferred regions in parameter space or for selected pairs of interesting observables by projecting the obtained points onto the corresponding plane, which corresponds to profiling over the parameters not plotted.

%%%%%%%%%%%%%%%%%%%%%%%%%%%%%%%%%%
\section{Scalar leptoquarks and $U(2)^5$ flavor symmetry}
\label{sec:LQandU2}

In the limit where only third generaton fermions are massive, the SM enjoys the global flavor symmetry \cite{Barbieri:2011ci,Barbieri:2012uh,Barbieri:2012tu}
\be
	G_F = U(2)_q \times U(2)_\ell \times U(2)_u \times U(2)_d \times U(2)_e~.
\ee
Masses of the first two generations of fermions and their mixing break this symmetry. In the quark sector the largest breaking is of size $\epsilon \approx y_t |V_{ts}| \approx 0.04$ \cite{Fuentes-Martin:2019mun}.
Formally, the symmetry breaking terms in the Yukawa matrices can be described in terms of spurions transforming under representations of $G_F$. The minimal set of spurions that can reproduce the observed masses and mixing angles is \footnote{Strictly speaking $V_\ell$ is not required in the SM, since in absence of neutrino masses lepton mixing is unphysical. It is however usually added for symmetry with the quark sector and, in our case, because it is needed in order to address the $R(K^{(*)})$ anomalies, which requires $|{\bf V}_\ell | \sim \mathcal{O}(0.1)$ \cite{Greljo:2015mma,Buttazzo:2017ixm}.}
\be\begin{split}
	& {\bf V}_q \sim ({\bf 2}, {\bf 1}, {\bf 1}, {\bf 1}, {\bf 1})~, \quad
	{\bf V}_\ell \sim ({\bf 1}, {\bf 2}, {\bf 1}, {\bf 1}, {\bf 1})~, \\
	& {\bf \Delta}_u \sim ({\bf 2}, {\bf 1}, {\bf \bar 2}, {\bf 1}, {\bf 1})~, \quad
	{\bf \Delta}_d \sim ({\bf 2}, {\bf 1}, {\bf 1}, {\bf \bar 2}, {\bf 1})~, \quad
	{\bf \Delta}_e \sim ({\bf 1}, {\bf 2}, {\bf 1}, {\bf 1}, {\bf \bar 2})~.
\end{split}\ee
In terms of these spurions the SM Yukawa matrices can be written as
\be
	Y_{u(d)} = y_{t(b)} \left( \begin{array}{c c}
				{\bf \Delta}_{u(d)} & x_{t(b)} {\bf V}_q \\
				0 & 1
			\end{array}\right)~, \qquad
	Y_{e} = y_\tau \left( \begin{array}{c c}
				{\bf \Delta}_{e} & x_\tau {\bf V}_\ell \\
				0 & 1
			\end{array}\right)~, \qquad
	\label{eq:U2YukSpurions}
\ee
with $x_{t,b,\tau}$ are $\mathcal{O}(1)$ complex numbers, ${\bf \Delta}$'s are $2\times 2$ matrices, and ${\bf V}_{q,\ell}$ are 2-component vectors.

In the context of the $B$-anomalies, this flavor symmetry was introduced as a possible explanation for the LFU breaking hints, that point to largest effects for $\tau$ leptons, smaller for muons, and even smaller for electrons. Furthermore, it was observed in Refs.~\cite{Greljo:2015mma,Barbieri:2015yvd,Bordone:2017anc,Buttazzo:2017ixm,Fuentes-Martin:2019mun,Marzocca:2018wcf,Gherardi:2020qhc} that the LQ couplings to second and third generations, required to fit the anomalies, were consistent with the expectations given by this symmetry. 
In this Section we study if, indeed, a complete implementation of $U(2)^5$ flavor symmetry for the $S_1$ and $S_3$ scalar LQs, including the couplings to first generation fermions, is consistent with the observed anomalies.

In the same flavor basis used to write the Yukawa couplings of Eq.~\eqref{eq:U2YukSpurions}, the $S_1$ and $S_3$ LQ couplings have the following structure:
\be
	\lambda^{1(3) L} = \lambda^{1(3)} \left( \begin{array}{c c}
				\tilde x^{1(3) L}_{q \ell} {\bf V}_q^* \times {\bf V}_\ell^\dagger  & \tilde x^{1(3) L}_{q} {\bf V}_q^* \\
				\tilde x^{1(3) L}_{\ell} {\bf V}_\ell^\dagger & \tilde x^{1(3) L}_{b\tau}
			\end{array}\right)~,
\ee
\be
	\lambda^{1 R} = \lambda^{1}_R \left( \begin{array}{c c}
				\mathcal{O}({\bf \Delta}_u {\bf V}_q {\bf \Delta}_e {\bf V}_\ell)  & \tilde x^{1 R}_{u} {\bf \Delta}_u^\dagger {\bf V}_q^* \\
				\tilde x^{1 R}_{e} {\bf V}_\ell^\dagger {\bf \Delta}_e^* & \tilde x^{1 R}_{t \tau}
			\end{array}\right) \approx \lambda^{1}_R \left( \begin{array}{c c}
				0  & 0 \\
				0 & \tilde x^{1 R}_{t \tau}
			\end{array}\right)~,
\ee
where $\lambda^{1(3)}$ and $\lambda^1_R$ are overall couplings, all $\tilde x$ are $\mathcal{O}(1)$ parameters, and in the last step in $\lambda^{1R}$ we neglected all the terms that give too small couplings to have a significant influence to our observables. In the following we can thus neglect the presence of the $\lambda^{1R}$ couplings in the $U(2)^5$ scenario since the $t_R \tau_R$ coupling does not affect in a relevant way the phenomenology.

By diagonalizing the SM lepton and down-quark Yukawa matrices one can put in relation some of the parameters in Eq.~\eqref{eq:U2YukSpurions} with observed masses and CKM elements, we refer to Ref.~\cite{Fuentes-Martin:2019mun} for a detailed discussion on this procedure.
For our purposes, the main result is that in this basis the quark doublet spurion is fixed by the CKM up to an overall $\mathcal{O}(1)$ factor, ${\bf V}_q = \kappa_q (V_{td}^*, V_{ts}^*)^T$, while the size of the leptonic doublet spurion $V_\ell \equiv |{\bf V}_\ell|$ as well as the angle that rotates left-handed electrons and muons, $s_e \equiv \sin \theta_e$, are free.
The same rotations that diagonalize the (lepton and down quark) Yukawas also apply to the LQ couplings.
The final result of this procedure is the following structure for the LQ couplings in the mass basis:
\be
	\lambda^{1(3) L} = \lambda^{1(3)} \left( \begin{array}{c c c}
			x^{1(3) }_{q \ell} s_e V_\ell  V_{td}  &  x^{1(3) }_{q \ell}  V_\ell  V_{td} & x^{1(3) }_{q}  V_{td}  \\
			x^{1(3) }_{q \ell} s_e V_\ell  V_{ts}  &  x^{1(3) }_{q \ell}  V_\ell  V_{ts}  & x^{1(3) }_{q}  V_{ts}  \\
			x^{1(3) }_{\ell} s_e V_\ell  &  x^{1(3) }_{\ell}  V_\ell  & 1 
			\end{array}\right)~.
	\label{eq:LQcouplU2}
\ee
All $ x^{1(3)}$ parameters are expected to be $\mathcal{O}(1)$ complex numbers and we absorbed the $\mathcal{O}(1)$ coefficient in the 3-3 component inside the definition of $\lambda^{1(3)}$.
It directly follows from Eq.~\eqref{eq:LQcouplU2} that the flavor symmetry imposes some strict relations between families:
\be
	\lambda^{1(3) L}_{1 \alpha} = \lambda^{1(3) L}_{2 \alpha} \frac{V_{td}}{V_{ts}}~, \qquad
	\lambda^{1(3) L}_{i 1} = \lambda^{1(3) L}_{i 2}  s_e ~. \qquad
	\label{eq:U2relations}
\ee
For the two LQs we thus remain with the two overall couplings $\lambda^{1(3)}$, that we can always take to be positive, six $\mathcal{O}(1)$ complex parameters ($ x^{1(3) }_{q \ell},  x^{1(3) }_{q},  x^{1(3) }_{\ell}$), one small angle $s_e$ that regulates the couplings to electrons compared to the muon ones, and finally $V_{\ell}$ that sets the size of muon couplings compared to tau ones. With the coupling structure of Eq.~\eqref{eq:LQcouplU2} the two leptoquarks decay dominantly to third generation fermions, since couplings to lighter generations are suppressed by the spurion factors.

Since $S_1$ does not mediate $d_i \to d_j \bar{\ell}_\alpha \ell_\beta$ at tree level, its contributions proportional to ${x}^1_\ell$ and ${x}^1_{q\ell}$ are only very weakly constrained. For this reason, to simplify the numerical scan we fix them to be equal to 1.\footnote{We checked that, as expected, if left free these parameters have an almost uniform distribution in the whole interval of interest $[-5, 5]$.}

We provide here some simplified expressions for the most relevant NP effects in this setup, deferring for details to App.~\ref{app:obs}:
\begin{eqnarray}
     \frac{\Delta R(D^{(*)})}{R(D^{(*)})_{\rm SM}} &\approx& 
        v^2 \left(1.09 \frac{ |\lambda^1|^2 (1 - x^{1 *}_q V^*_{tb} )}{2 M_1^2} - 1.02 \frac{ |\lambda^3|^2 (1 - x^{3 *}_q V^*_{tb} )}{2 M_3^2} \right)~, \\
	\Delta C_9^{sb\mu\mu} = - \Delta C_{10}^{sb\mu\mu} &\approx& \frac{\pi}{\sqrt{2} G_F \alpha V_{tb}} \frac{|\lambda^3|^2 |V_\ell |^2 x^3_\ell x^{3*}_{q\ell}}{M_3^2}~, \\
	\Delta C_9^{sd\mu\mu} = - \Delta C_{10}^{sd\mu\mu} &\approx& \frac{\pi V_{ts}^* V_{td}}{\sqrt{2} G_F \alpha} \frac{|\lambda^3|^2 |V_\ell |^2 | x^{3}_{q\ell}|^2 }{M_3^2}~, \label{eq:exprKLmumu}\\
	\left[ L^{VLL}_{\nu d} \right]_{\nu_\tau \nu_\tau s b} &\approx&
		V_{ts}^* \left( \frac{|\lambda^1|^2 x^{1 *}_q}{2 M_1^2} + \frac{ |\lambda^3|^2 x^{3 *}_q }{2 M_3^2} \right)~, \\
	\left[ L^{VLL}_{\nu d} \right]_{\nu_\tau \nu_\tau d s} &\approx&
		V_{td}^* V_{ts} \left( \frac{|\lambda^1|^2 |x^1_q|^2}{2 M_1^2} + \frac{|\lambda^3|^2 |x^3_q|^2}{2 M_3^2} \right)~, \label{eq:exprKpinunu} \\
   10^3 \delta g^Z_{\tau_L} &\approx& 0.59 \frac{|\lambda^1|^2}{M_1^2 / \TeV^2} + 0.80 \frac{|\lambda^1|^2}{M_1^2 / \TeV^2} ~, \label{eq:exprZtautau}
\end{eqnarray}
\be
    C^1_K \approx \frac{V_{ts}^* V_{td}}{128 \pi^2} \left( \frac{|\lambda^1|^4 |x^1_q|^4}{M_3^2} + 5 \frac{|\lambda^3|^4 |x^3_q|^4}{M_3^2} + \frac{|\lambda^1|^2 |\lambda^3|^2 |x^1_q|^2|x^3_q|^2 \log M_3^2/M_1^2}{M_3^2 - M_1^2}\right)~,  \label{eq:exprC1K}
\ee
where $\left[ L^{VLL}_{\nu d} \right]_{\nu_\tau \nu_\tau d_i d_j}$ are the Wilson coefficients (WCs) of the low-energy operators\\ \mbox{$(\bar\nu_\tau \gamma_\mu \nu_\tau)(\bar{d}^i_L \gamma^\mu d^j_L)$}, $\delta g^Z_{\tau_L}$ describes the deviation in the $Z$ couplings to $\tau_L$, and $C^1_K$ is the coefficient of the $(\bar s \gamma_\mu P_L d)^2$ operator.
The leading contribution to $s \to d \mu \mu$ transitions has a phase fixed to be equal to the SM one, so no large effect in $K_S \to \mu \mu$ can be expected.
Analogously, also in $s \to d \nu\nu$ the NP coefficients have the same phase as in the SM, since the $x$ coefficients enter with the absolute value squared. This implies that no cancellation between the two LQs can take place in this channel and that we expect a tight correlation between $K_L \to \pi^0 \nu \nu$ and $K^+ \to \pi^+ \nu\nu$, independently on the phases of the couplings.
Similar considerations apply for all $s \to d$ transitions. On the other hand, non-trivial phases can appear in $b \to s$ transitions and a mild cancellation can alleviate the $B\to K^{(*)}\nu\nu$ bound \cite{Buttazzo:2017ixm}.
Since real couplings are favored by the $B$-anomalies and since in any case the phases in Kaon physics observables are fixed by the $U(2)^5$ flavor structure, in our numerical analysis we only consider real values for all parameters.

\begin{figure}[t]
\centering
\includegraphics[height=6cm]{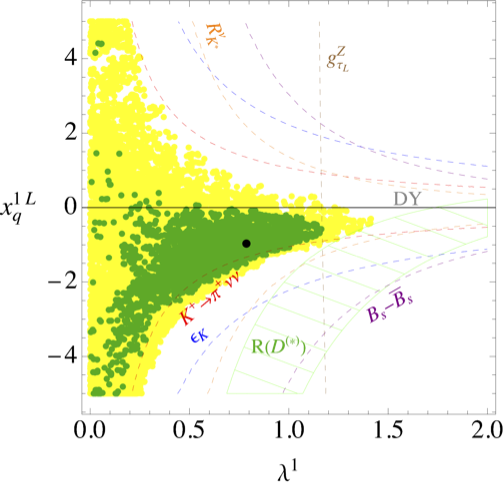} \quad
\includegraphics[height=6cm]{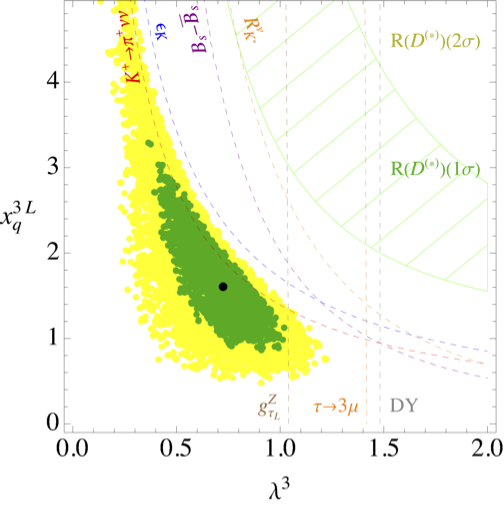} \\[0.5cm]
\includegraphics[height=6cm]{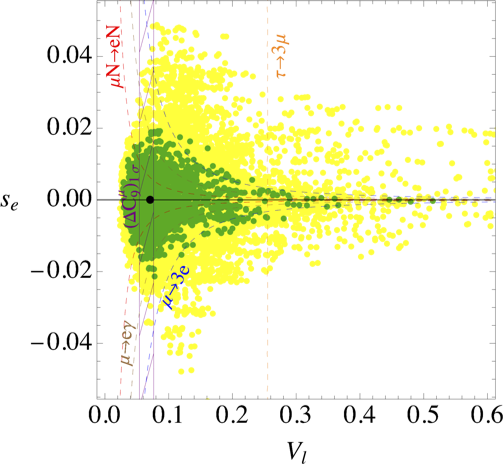} \quad
\includegraphics[height=6cm]{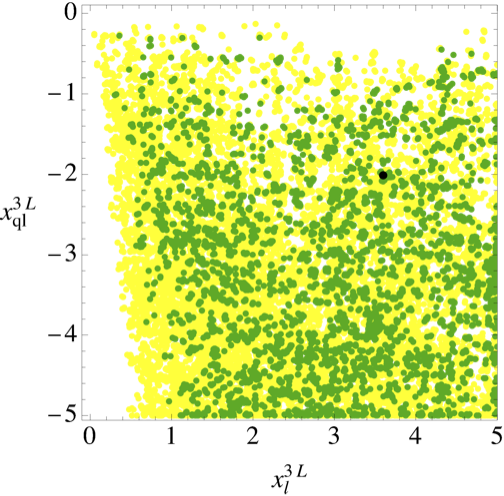} \\
\caption{\small\label{fig:U2_params} Results of a parameter scan in the $U(2)^5$ scenario. The green (yellow) points are within the 68\% (95\%) CL from the best-fit point (shown in black). We also overlay 2$\sigma$ constraints from single observables, where other parameters are fixed to the best-fit point \eqref{eq:U2bestfit}.}
\end{figure}

\subsection{Analysis and discussion}

Using the global likelihood we find the best-fit point in  parameter space, allowing the $x$'s to vary in the range $|x|<5$, while $\lambda^{1(3)}, V_\ell > 0$. Fixing $M_1 = M_3 = 1.1\TeV$ we get $\chi^2_{\SM} - \chi^2_{\rm best-fit} = 47.6$, for:\footnote{This puts the LQs above present LHC limits, see e.g. Ref.~\cite{Saad:2020ihm} for a recent review. We note that the fit slightly worsens when the masses are increased.}
\be
    \text{best-fit } U(2)^5: \quad
    \begin{array}{l l l l}
		\lambda^{1} \approx 0.79~, &
		\lambda^{3} \approx 0.72~, &
		V_\ell \approx 0.071~, &
		s_e \approx 0~, \\
		x^1_q \approx -0.97~,	&	
		x^3_q \approx 1.6~, &
		x^3_{\ell} \approx 3.6~, &
		x^3_{q\ell} \approx -2.0~. 
	\end{array}
\label{eq:U2bestfit}
\ee
We then perform a numerical scan on all the parameters in Eq.~\eqref{eq:U2bestfit}, selecting only points with a $\Delta \chi^2 = \chi^2 - \chi^2_{\rm best-fit}$ corresponding to a 68\% (green points) or 95\% (yellow) confidence level.
The results are shown in Fig.~\ref{fig:U2_params}, where we project the points to several 2D planes. We also show $2\sigma$ constraints from single observables, obtained by fixing the parameters not in the plot to the corresponding best-fit values, Eq.~\eqref{eq:U2bestfit}.\footnote{This procedure misses the correlations between the plotted parameters and those that are fixed, that is however fully kept in the parameter scan. For this reason the bounds shown in this way are useful mainly to gain an understanding of the relevant observables in each plane.}
We observe from Fig.~\ref{fig:U2_params} (bottom-left) that values $V_\ell \approx 0.1$ and $|s_e| \lesssim 0.02$ are preferred.
We can also see that all the $x$'s can be of $\mathcal{O}(1)$, with no tendency towards parametrically smaller or larger values, hence the structure of the $U(2)^5$ symmetry is respected.

\begin{figure}[t]
\centering
\includegraphics[height=6cm]{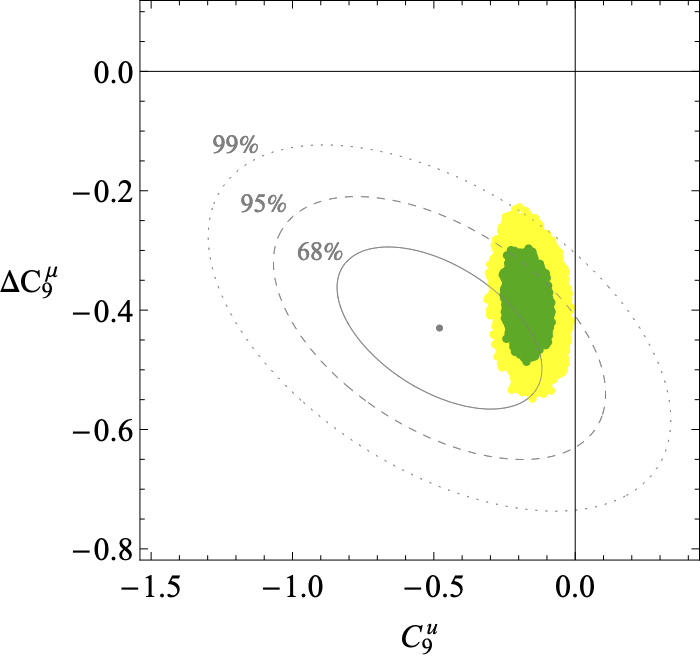} \quad
\includegraphics[height=6cm]{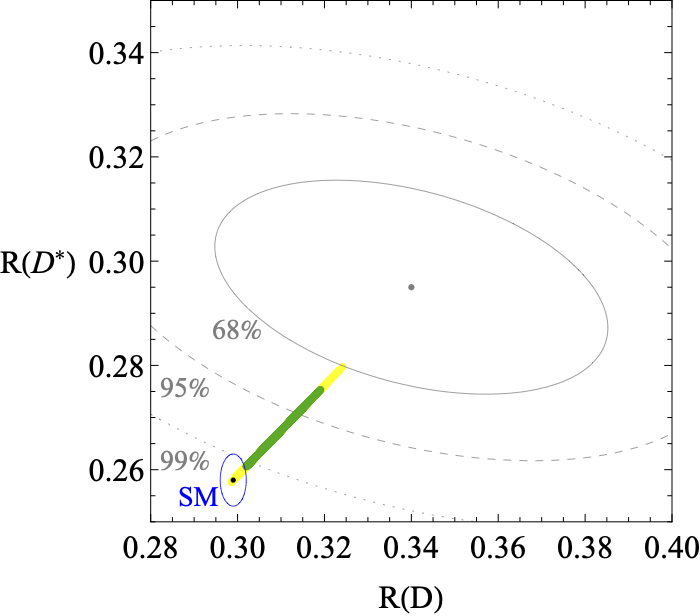} \\[0.5cm] 
\caption{\small\label{fig:U2_obs_B} Results, for the $B$-anomalies, of the parameter scan in the $U(2)^5$ scenario shown in Fig.~\ref{fig:U2_params}. The green (yellow) points are within the 68\% (95\%) CL from the best-fit point \eqref{eq:U2bestfit}. The gray lines describe different CL regions in the fit of the anomalies  (see App.~\ref{app:obs} for details).}
\end{figure}

\begin{figure}[p]
\centering 
\includegraphics[height=8cm]{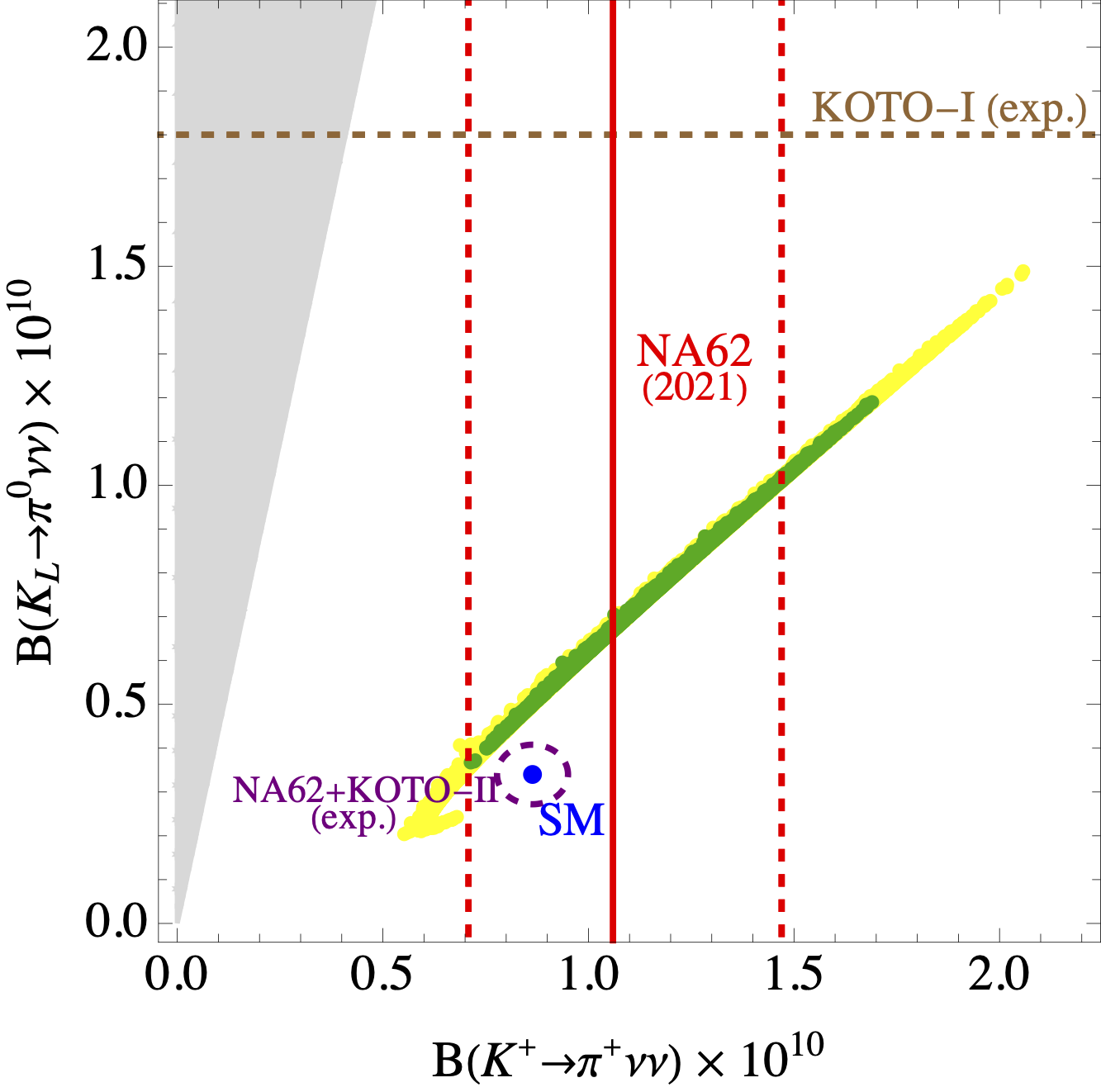} \\[0.5cm] 
\includegraphics[height=7cm]{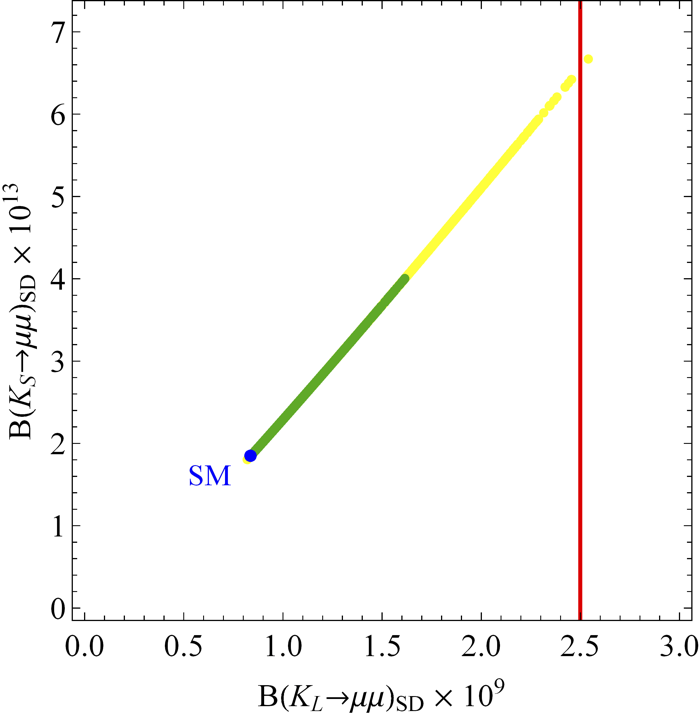} \quad
\includegraphics[height=7cm]{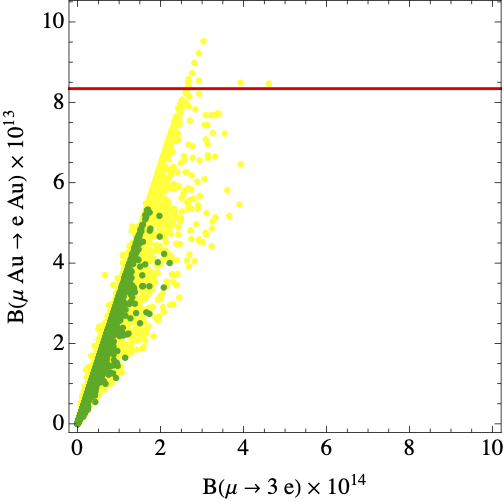} \\
\caption{\small\label{fig:U2_obs_K} Results, for rare Kaon decays and $\mu \to e$ conversion processes, of the parameter scan in the $U(2)^5$ scenario shown in Fig.~\ref{fig:U2_params}. The green (yellow) points are within the 68\% (95\%) CL from the best-fit point. In the upper plot, the gray region is excluded by the Grossman-Nir bound~\cite{Grossman:1997sk}, the red solid and dashed lines represent the present measurement from NA62, the dotted brown line the sensitivity prospect for KOTO after stage-I, while the dotted purple one the final sensitivity expected from NA62 and KOTO (stage-II).
In the lower plot the red lines describe the present 95\% CL bound (see App.~\ref{app:obs} for details).}
\end{figure}

In Figs.~\ref{fig:U2_obs_B} and \ref{fig:U2_obs_K} we show the values of particularly interesting pairs of observables obtained with the same sets of parameter-space points.
From Fig.~\ref{fig:U2_obs_B} we observe that, while neutral-current $B$-anomalies can be addressed entirely, this setup can enhance $R(D^{(*)})$ only by $\lesssim 7\%$ of the SM size, i.e. at the $2\sigma$ level of the present combination.
This situation should be compared with the result of the analogous similar fit with $S_1$ and $S_3$ with only couplings to left-handed fermions shown in \cite{Gherardi:2020qhc} (c.f. Fig.~5), where both anomalies can be satisfied in a scenario where couplings to the second generation quarks were compatible with a $U(2)^5$ flavor structure, $|\lambda^{1(3)L}_{s \ell}| \sim |V_{ts}/V_{tb}| |\lambda^{1(3)L}_{b \ell}|$, but couplings to first generation were set to zero.

Therefore, the reason for the inability of the $U(2)^5$-symmetric scenario to fully address charged-current anomalies must be found in first-generation constraints, specifically Kaon physics.
Indeed, this can be seen in the first row of Fig.~\ref{fig:U2_params}, where we observe that the bounds from $K^+ \to \pi^+ \bar\nu \nu$, Eq.~\eqref{eq:exprKpinunu}, and $\epsilon_K$ (i.e. $\Im{C^1_K}$), Eq.~\eqref{eq:exprC1K}, in combination with the constraints on $\lambda^{1,3}$ from $Z\to \bar\tau\tau$, Eq.~\eqref{eq:exprZtautau}, don't allow the fit to enter the region preferred by $R(D^{(*)})$, due to the precise relations between couplings to the first and the second generation, derived from the flavor structure, i.e. Eq.~\eqref{eq:U2relations}.

Regarding Kaon physics observables, from Fig.~\ref{fig:U2_obs_K} we see that $\Br(K^+ \to \pi^+ \bar\nu \nu)$ can take all values currently allowed by the NA62 bound \cite{NA62:2021zjw} (we show with vertical lines the best-fit and the $\pm1\sigma$ intervals) and therefore any future update on this observable will put further strong constraints on this scenario. Furthermore, since the phase in $s \to d \nu \nu$ is fixed by the corresponding CKM phase, Eq.~\eqref{eq:exprKpinunu}, a correlation between this mode and $\Br(K_L \to \pi^0 \bar\nu \nu)$ is obtained, with values $\sim 10^{-10}$ also for the latter.\footnote{The same correlation takes place also in Minimal Flavor Violation setups, where also the phase of the relevant Wilson coefficient is fixed, as was shown in Ref.~\cite{Buras:2001af}.}
Therefore, even by the end of stage-I the KOTO experiment won't be able to reach the sensitivity to test this model (brown horizontal dotted line).
However, the future sensitivity goals by NA62 ($10\%$ \cite{Ruggiero:2017hjh}) and KOTO stage-II  , or KLEVER, ($20\%$ \cite{Aoki:2021cqa,Ambrosino:2019qvz}) would be able to completely test this scenario (purple ellipse). 

The model also predicts short-distance contributions to $\Br(K_L\to \mu\mu)$ of the order of the present bound, although it remains challenging to improve this constraint in the future due to non-perturbative contributions to the long-distance component. Also in this channel the phase of the NP WC is fixed to the one of $V_{ts}^* V_{td}$, see Eq.~\eqref{eq:exprKLmumu}. The $K_S \to \mu\mu$ mode is thus completely correlated with the $K_L$ decay and the New Physics effect adds constructively to the SM short-distance amplitude, the expected size is however below the SM long-distance contribution of $\approx 5\times 10^{-12}$.
The expected relative effect in the tree-level transitions $s \to u \mu \bar \nu_\mu$ and $d \to u \mu \bar \nu_\mu$ is of order $10^{-7}$ and $10^{-9}$, respectively, excluding possible signatures from these decays.

Finally, the rotation angle between electrons and muons, $s_e$, is constrained mainly by LFV $\mu \to e$ processes, the strongest bound presently given by $\mu \to e$ conversion in gold atoms, while the predicted effect in titanium is strictly correlated in our model, with an approximate relation $\cB(\mu\to e)_{\rm Au} \approx 1.3 \cB(\mu\to e)_{\rm Ti}$.
In Fig.~\ref{fig:U2_obs_K} we show the correlation with $\mu \to 3 e$, while the expected branching ratio for $\mu \to e \gamma$ is $\lesssim 0.5 \times 10^{-13}$, thus below than the present precision.
These measurements are expected to improve substantially in the future, reaching limits of the order of $10^{-16}$ in $\mu \to e$ conversion in nuclei from COMET and Mu2e \cite{Kuno:2013mha,Ankenbrandt:2006zu,Knoepfel:2013ouy,Bartoszek:2014mya} or even $10^{-18}$ by the PRISM proposal, and also a level of $10^{-16}$ in $\mu \to 3 e$ from the Mu3e experiment \cite{Mu3e:2020gyw}. These will put further constraints on the $s_e$ parameter, or a signal could be observed if this mixing angle is large enough.

For what regards LFV in Kaon decays, given the largest allowed values for $s_e$ and $V_\ell$ from present limits, we obtain at most $\Br(K_L \to \mu e) \lesssim 10^{-15}$ and $\Br(K^+ \to \pi^+ \mu e) \lesssim 10^{-18}$.

%%%%%%%%%%%%%%%%%%%%%%%%%%%%%%%%%%
\section{General case with right-handed couplings}
\label{sec:LQgeneral}

In this section we depart from the flavor symmetry assumption and examine the fit of the $S_1 + S_3$ model in the general case where all the couplings, including the right-handed ones, are allowed.
As investigated in Ref.~\cite{Gherardi:2020qhc}, if the couplings are a priori uncorrelated, there is enough freedom to accommodate both $B$-physics anomalies as well as the discrepancy in the anomalous magnetic moment of the muon.

Due to the high dimensionality of the parameter space, it is computationally too expensive to perform a $\chi^2$ minimization by random search techniques. For the purposes of our discussion it suffices then to use the best-fit point of Eq. (3.11) in Ref.~\cite{Gherardi:2020qhc}\footnote{A fit with the updated values for the various flavor constraints including the new measurement of $R_K$ does not result in any substantial variations.} in order to fix the relevant couplings and let only the additional couplings vary.
A further simplification constitutes in switching off the couplings to electrons. This choice is justified by the fact that the necessary suppression required to pass the stringent bounds from LFV $\mu \to e$ processes is of the same order in both the flavor-symmetry motivated case and the general one.

\begin{figure}[t]
\centering
\includegraphics[height=7cm]{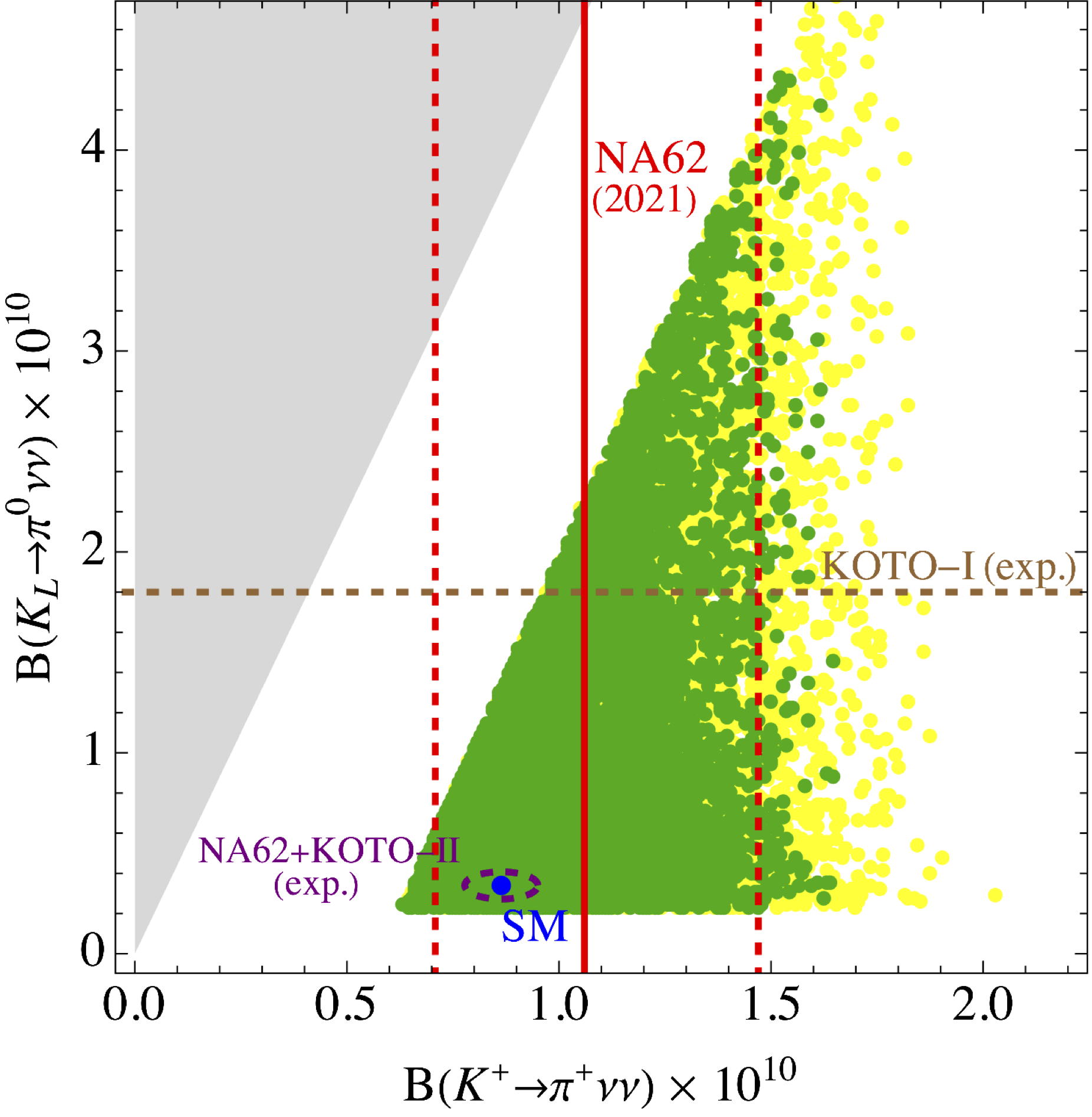} \quad
\includegraphics[height=7cm]{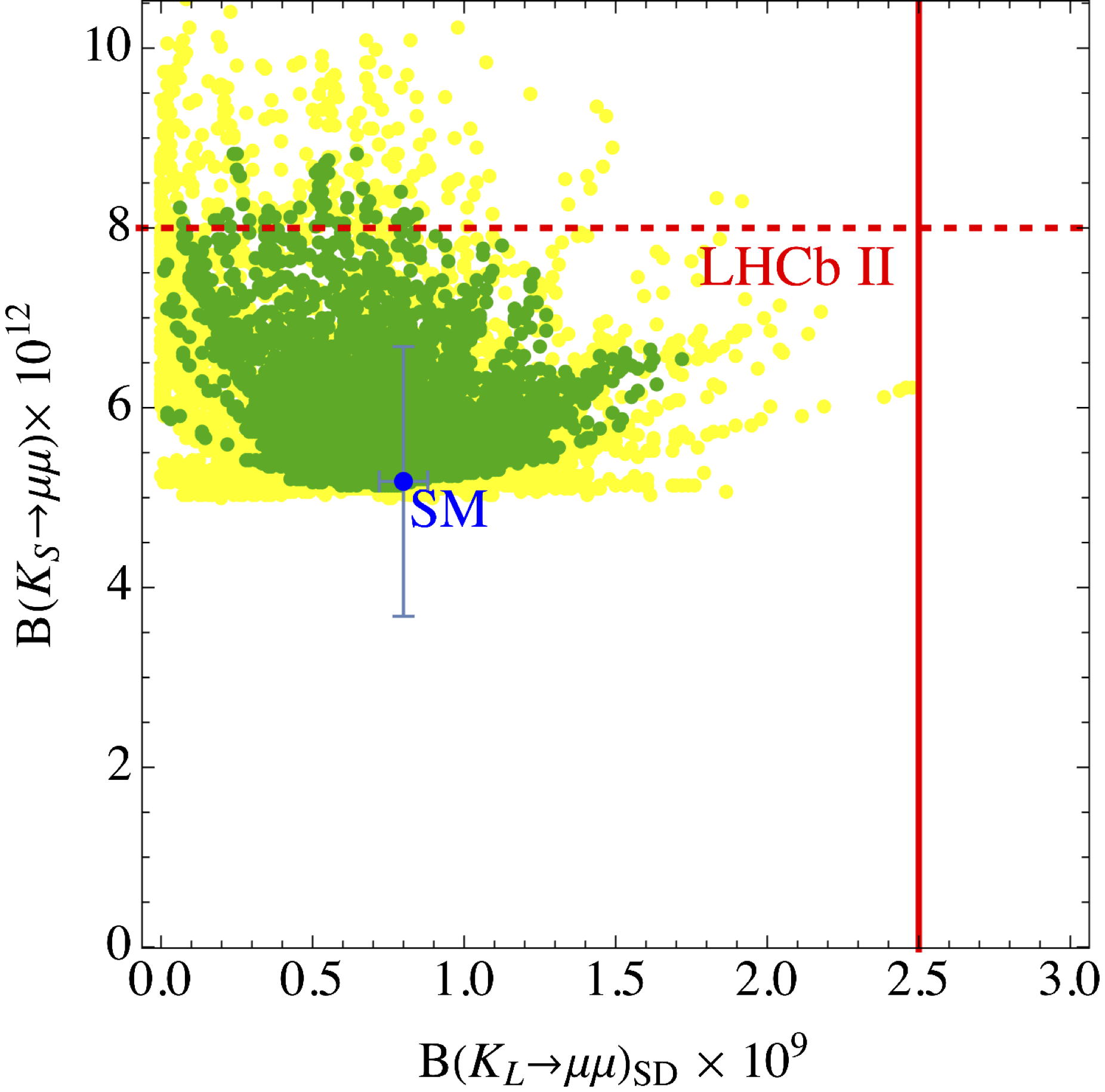} 
\caption{\label{fig:K_decays_general} Allowed region obtained by varying the couplings $\lambda_{s\tau}^{1(3)L}$ and $\lambda_{d\tau}^{1(3)L}$,relevant for the $B \to K \nu\nu$ decays (left), and the couplings $\lambda_{s\mu}^{3L}$ and $\lambda_{d\mu}^{3L}$ relevant for the $K_{L,S} \to \mu\mu$ decays (right). The rest of the couplings are fixed to the best-fit point couplings in Eq. (3.11) of Ref.~\cite{Gherardi:2020qhc} and compatibility with the global fit is retained at 68\% (green) and 95\% (yellow) CL. On the left plot, the gray region is excluded by the Grossman-Nir bound, the red solid and dashed lines represent the present measurement from NA62, the dotted brown line the sensitivity prospect for KOTO after stage-I, while the dotted purple one the final sensitivity expected from NA62 and KOTO (stage-II). On the right plot, the red solid line represents the bound set in Ref.~\cite{Isidori:2003ts} and the red dashed line the future prospects by LHCb \cite{Bediaga:2018lhg}.}
\end{figure}

Using the best-fit point from Ref.~ \cite{Gherardi:2020qhc} assures that all the anomalies are addressed within $1\sigma$ (c.f. Fig.~5 therein).
In order to assess the viable values of $\Br(K_L \to \pi^0 \nu\nu)$ and $\Br(K^+ \to \pi^+ \nu\nu)$ in this setup we perform a likelihood scan of the complex couplings $\lambda_{s\tau}^{1(3)L}$ and $\lambda_{d\tau}^{1(3)L}$  (see Eq.~\eqref{eq:Kpinunu}) and keep only values that have a likelihood within the 68\%  or 95\% CL from the best-fit point. The corresponding branching ratios are reported in the left plot of Fig.~\ref{fig:K_decays_general}, from which we observe that, compared to the $U(2)^5$ case, the viable values are significantly expanded.
We also notice that the two decays are induced by the same short distance operators and are trivially related through isospin, leading to the so-called Grossman-Nir bound~\cite{Grossman:1997sk}. The LQ interactions fall under a category of models that may saturate the bound~\cite{Grossman:2003rw}. Yet this is not entirely possible due to the constraints on the magnitude of the couplings to muons and electrons. 
Nevertheless, we see that $\Br(K_L \to \pi^0 \nu \nu)$ could potentially take values that can be probed by the end of stage 1 of the KOTO experiment. For other studies of correlations between the $K^+$ and the $K_L$ modes see e.g. Refs~\cite{Buras:2001af,Blanke:2009pq,Buras:2015yca,Bordone:2017lsy}.

By letting vary the $\lambda_{s\mu}^{3L}$ and $\lambda_{d\mu}^{3L}$ complex couplings we find that the short distance contribution to $K_{S} \to \mu\mu$ can reach values of the order of the long-distance one, while that to the $K_L$ mode saturate the theory-derived constraint~\cite{Isidori:2003ts}. 
In the right plot of Fig.~\ref{fig:K_decays_general} we present the 68\%  and 95\% CL regions from the best-fit point in these two observables.
Considering the $\approx 30\%$ uncertainty on the SM prediction for the $K_S$ mode,
we notice that a significant part of the preferred region features NP effects that are distinguishable from the SM ones. As a matter of fact, the future prospects look promising, since the LHCb Upgrade II plans to exclude branching fractions down to near the SM prediction~\cite{Bediaga:2018lhg} (see dashed line in Fig. \ref{fig:K_decays_general}). \par
Regarding $\mu \to e$ transition in nuclei we expect a very similar result to the one shown in Fig.~\ref{fig:U2_obs_K}, since even in that case the observables saturate the present bounds and there is additional freedom if the flavor symmetry assumption is removed.

%%%%%%%%%%%%%%%%%%%%%%%%%%%%%%%%%%
\section{Conclusions}
\label{sec:conclusions}

The observed anomalies in $B$ decays can potentially be addressed in LQ scenarios. While couplings to first generation of fermions are not required to describe these deviations, and could therefore be set to zero in a bottom-up approach, the typical expectation from UV models is that all couplings should be generated.
In this work we study the impact that LQ couplings to first generation fermions can have on Kaon and electron LFV observables, in scenarios that aim at addressing the $B$-anomalies.

Correlations between couplings to different generations, and therefore between $B$ and Kaon decays, can be obtained only if the flavor structure is specified. In this work we consider the approximate $U(2)^5$ flavor symmetry, that is motivated from the observed SM fermions mass hierarchies and also predicts a pattern of deviations in $B$ decays consistent with the observed one.
With this assumption, the LQ coupling to left-handed $d$ quarks are fixed to be equal to those to left-handed $s$ quarks times the small CKM factor $V_{td}/V_{ts}$ (see Eq.~\eqref{eq:U2relations}). This structure correlates strongly Kaon physics with $B$ decays. After performing a global likelihood analysis of a large set of observables in this scenario, we find that the $S_1 + S_3$ LQs can address the $R(D^{(*)})$ anomalies only at the $2\sigma$ level.
The most important observables preventing a successful fit are $\Br(K^+ \to \pi^+ \nu\nu)$, $\epsilon_K$, and the $Z$ couplings to $\tau$.
We thus expect future improved measurements of $K^+ \to \pi^+ \nu\nu$ from NA62 to further test this scenario.
The LQ couplings to electrons, instead, are constrained mainly by the limits from $\mu \to e$ transition in nuclei and $\mu \to 3 e$ decay, that will be improved by several orders of magnitude in the near future.

Going beyond the flavor symmetric scenario, we also studied the allowed values for $\Br(K^+ \to \pi^+ \nu\nu)$, $\Br(K_L \to \pi^0 \nu\nu)$, and $K_{L/S} \to \mu^+ \mu^-$ in the general case where no flavor structure is imposed. We find that, in this setup, values of $\Br(K_L \to \pi^0 \nu\nu)$ and $\Br(K_S \to  \mu^+ \mu^-)$ that could be potentially probed by KOTO stage-1 or LHCb, respectively, are allowed.

The $B$-anomalies are expected to receive further experimental inputs in the next few years by LHCb, Belle-II, as well as CMS and ATLAS experiments. However, even assuming these will be confirmed, in order to understand the flavor structure of the underlying New Physics the connection to Kaon physics and to observables sensitive to electron couplings will be crucial.

%%%%%%%%%%%%%%%%%
\subsection*{Acknowledgements}
%%%%%%%%%%%%%%%%%

DM and ST acknowledge support by MIUR grant PRIN 2017L5W2PT. DM is also partially supported by the INFN grant SESAMO and by the European Research Council (ERC) under the European Union’s Horizon 2020 research and innovation programme, grant agreement 833280 (FLAY). EV has been partially funded by the Deutsche Forschungsgemeinschaft (DFG, German
Research Foundation) under Germany's Excellence Strategy - EXC-2094 - 390783311,
by the Collaborative Research Center SFB1258 and the BMBF grant 05H18WOCA1 and
thanks the Munich Institute for Astro- and Particle Physics (MIAPP) for hospitality.

%%%%%%%%%%%%%%%%%%%%%%%%%%%%%%%%%%%%%
%%%%%%%%%%%%%%%%%%%%%%%%%%%%%%%%%%%%%
%%%%%%%%%%%%%%%%%%%%%%%%%%%%%%%%%%%%%

\begin{table}[p]
\begin{center}
\begin{tabular}{ | c | c | c | }
\hline
\textbf{Observable} & \textbf{SM prediction} & \textbf{Experimental bounds} \\
\hline
\hline
\cellcolor[gray]{0.92} $b\to s \ell \ell$ observables & \cellcolor[gray]{0.92} & \cellcolor[gray]{0.92} \cite{Gherardi:2020qhc}\\
\hline
$\Delta\mathcal{C}_{9}^{sb\mu\mu}$ & 0 &  $-0.43\pm 0.09$ \cite{Aebischer:2019mlg} \\
\hline
$\mathcal{C}_{9}^{\text{univ}}$ & 0 &  $-0.48\pm 0.24$ \cite{Aebischer:2019mlg} \\
\hline
\hline
\cellcolor[gray]{0.92} $b\to c \tau(\ell) \nu$ observables & \cellcolor[gray]{0.92}  &  \cellcolor[gray]{0.92} \cite{Gherardi:2020qhc}\\
\hline
$R_D$ & $0.299\pm 0.003$ \cite{Amhis:2016xyh} & $0.34\pm 0.027\pm 0.013$ \cite{Amhis:2016xyh} \\
\hline
 $R_D^*$ & $0.258\pm 0.005$ \cite{Amhis:2016xyh} & $0.295\pm 0.011\pm 0.008$ \cite{Amhis:2016xyh} \\
 \hline   
 $P_\tau^{D^*}$ & $-0.488\pm 0.018$ \cite{Bordone:2019vic} & $-0.38\pm 0.51\pm 0.2\pm 0.018$ \cite{Hirose:2017dxl} \\
 \hline   
 $F_L$ & $0.470\pm 0.012$ \cite{Bordone:2019vic} & $0.60\pm 0.08\pm 0.038\pm 0.012$ \cite{Abdesselam:2019wbt} \\
 \hline
 $\Br(B_c^+\to \tau ^+ \nu)$ & $2.3 \%$ & $< 10\%$ (95\% CL) \cite{Akeroyd:2017mhr}  \\   
 \hline
	$R_D^{\mu/e}$ & 1 &  $0.978 \pm 0.035$ \cite{Aubert:2008yv,Glattauer:2015teq} \\
 \hline   
 \hline
 \cellcolor[gray]{0.92} $b\to s \nu\nu$ and $s\to d \nu\nu$ & \cellcolor[gray]{0.92} &  \cellcolor[gray]{0.92} \cite{Gherardi:2020qhc}\\
 \hline
 $R_K^\nu$ & 1 \cite{Buras:2014fpa} & $<  4.7$  ~\cite{Grygier:2017tzo} \\
 \hline
  $R_{K^*}^\nu$ & 1 \cite{Buras:2014fpa} & $<  3.2$   ~\cite{Grygier:2017tzo}  \\
 \hline
 \hline
 \cellcolor[gray]{0.92} $b\to d \mu\mu$ and $b\to d ee$ & \cellcolor[gray]{0.92} &  \cellcolor[gray]{0.92} App.~\ref{app:bdmumu}\\
 \hline
 $\Br(B^0\to \mu\mu)$ & $(1.06 \pm 0.09) \times 10^{-10}$ ~\cite{Bobeth:2013uxa, Bailey:2015nbd} & $(1.1 \pm 1.4) \times 10^{-10}$  ~\cite{Chatrchyan:2013bka,Aaij:2015nea} \\
 \hline
  $\Br(B^+\to \pi^+ \mu\mu)$ & $(2.04\pm 0.21) \times 10^{-8}$ ~\cite{Bobeth:2013uxa, Bailey:2015nbd} & $(1.83\pm 0.24) \times 10^{-8}$   ~\cite{Chatrchyan:2013bka,Aaij:2015nea}  \\
 \hline  
$\Br(B^0\to ee)$ & $(2.48\pm 0.21) \times 10^{-15}$~\cite{Bobeth:2013uxa, Bailey:2015nbd} & $<  8.3 \times 10^{-8}$ ~\cite{Tanabashi:2018oca} \\
 \hline
$\Br(B^+\to \pi^+ ee)$ & $(2.04\pm 0.24) \times 10^{-8} $ ~\cite{Bobeth:2013uxa, Bailey:2015nbd}& $<  8 \times 10^{-8}$  ~\cite{Tanabashi:2018oca} \\  
\hline
\hline
\cellcolor[gray]{0.92} $B$ LFV decays & \cellcolor[gray]{0.92} &  \cellcolor[gray]{0.92} \cite{Gherardi:2020qhc}\\       
\hline
$\Br(B_d\to \tau^\pm \mu^\mp)$ & 0  &$< 1.4 \times 10^{-5}$ ~\cite{Aaij:2019okb}  \\      
\hline 
$\Br(B_s\to \tau^\pm \mu^\mp)$ &  0 &$< 4.2 \times 10^{-5}$ ~\cite{Aaij:2019okb}  \\      
\hline  
 $\Br(B^+\to K^+ \tau^- \mu^+)$ & 0  &$< 5.4 \times 10^{-5}$ ~\cite{Lees:2012zz}  \\ 
\hline
  \multirow{2}{*}{$\Br(B^+\to K^+ \tau^+ \mu^-)$} &  \multirow{2}{*}{0} &$< 3.3 \times 10^{-5}$ ~\cite{Lees:2012zz}  \\     
 & & $< 4.5 \times 10^{-5}$ ~\cite{Aaij:2020mqb} \\   
\hline 
 \end{tabular}
\caption{ Observables from $B$ and $D$ meson decays. Upper limits correspond to 95\%CL. \label{tab:obs}}
\end{center} 
\end{table}

\begin{table}[p]
\begin{center}
\begin{tabular}{ | c | c | c | }
\hline
\textbf{Observable} & \textbf{SM prediction} & \textbf{Experimental bounds} \\
 \hline  
\hline
\cellcolor[gray]{0.92} $D$ leptonic decay & \cellcolor[gray]{0.92} & \cellcolor[gray]{0.92} \cite{Gherardi:2020qhc} and App.~\ref{app:rareD}\\       
\hline
$\Br(D_s\to \tau\nu)$ & $(5.169\pm 0.004)\times 10^{-2}$~\cite{Aoki:2016frl}  &$(5.48 \pm 0.23) \times 10^{-2}$~\cite{Tanabashi:2018oca} \\   
\hline
 $\Br(D^0 \to \mu\mu)$ & $\approx 10^{-11}$ \cite{Gisbert:2020vjx} & $< 7.6 \times 10^{-9}$~\cite{Aaij:2013cza} \\
\hline
 $\Br(D^+ \to \pi^+ \mu\mu)$ & $ \mathcal{O}(10^{-12})$ \cite{deBoer:2015boa} & $< 7.4 \times 10^{-8}$~\cite{Aaij:2020wyk} \\
\hline
\hline
 \cellcolor[gray]{0.92} Rare Kaon decays ($\nu\nu$) & \cellcolor[gray]{0.92} &  \cellcolor[gray]{0.92} App.~\ref{sec:Kpinunu}\\
 \hline
$  \Br(K^+ \to \pi^+ \nu\nu)   $ & $8.64 \times 10^{-11}$~\cite{Buras:2015yca} & $ (11.0\pm 4.0) \times 10^{-11}$ \cite{CortinaGil:2020vlo} \\
 \hline
$ \Br(K_L \to \pi^0 \nu\nu) $ & $3.4 \times 10^{-11} $ ~\cite{Buras:2015yca}& $< 3.6 \times 10^{-9}$  \cite{Ahn:2018mvc} \\  
\hline
 \hline
 \cellcolor[gray]{0.92} Rare Kaon decays ($\ell\ell$) & \cellcolor[gray]{0.92} &  \cellcolor[gray]{0.92} App.~\ref{app:KLmue} and~\ref{app:Kll}\\
 \hline
 $\Br(K_L\to \mu\mu)_{SD}$ & $8.4\times 10^{-10}$~\cite{Buchalla:1993wq} & $< 2.5 \times 10^{-9}$~\cite{Isidori:2003ts} \\
 \hline
  $\Br(K_S\to \mu\mu)$ & $(5.18 \pm 1.5)\times 10^{-12}$~\cite{Ecker:1991ru,Isidori:2003ts,DAmbrosio:2017klp} & $< 2.5 \times 10^{-10}$ ~\cite{Aaij:2020sbt} \\
 \hline  
$\Br(K_L\to \pi^0 \mu\mu)$ & $(1.5\pm 0.3)\times 10^{-11}$~\cite{Isidori:2004rb} & $< 4.5 \times 10^{-10}$~\cite{AlaviHarati:2000hs} \\
 \hline
$\Br(K_L\to \pi^0 ee)$ & $(3.2^{+ 1.2}_{-0.8})\times 10^{-11}$~\cite{Buchalla:2003sj} & $< 2.8 \times 10^{-10}$~\cite{AlaviHarati:2003mr} \\  
\hline
 \hline
 \cellcolor[gray]{0.92} LFV in Kaon decays & \cellcolor[gray]{0.92} &  \cellcolor[gray]{0.92} App.~\ref{app:KLmue} and~\ref{app:Kll}\\
\hline
$\Br(K_L\to \mu e)$ & 0 & $< 4.7 \times 10^{-12}$~\cite{Ambrose:1998us} \\
\hline
$\Br(K^+\to \pi^+ \mu^- e^+)$ & 0 & $< 7.9 \times 10^{-11}$~\cite{CortinaGil:2021mkg} \\
\hline
$\Br(K^+\to \pi^+ e^- \mu^+)$ & 0 & $< 1.5 \times 10^{-11}$~\cite{Abouzaid:2007aa} \\
 \hline       
 \hline
\cellcolor[gray]{0.92} CP-violation & \cellcolor[gray]{0.92} &\cellcolor[gray]{0.92} App.~\ref{app:epsprime} \\
\hline
$\epsilon_K'/\epsilon_K$ & $(15 \pm 7) \times 10^{-4}$~\cite{Gisbert:2018tuf} & $(16.6 \pm 2.3) \times 10^{-4}$~\cite{Tanabashi:2018oca} \\
\hline
 \end{tabular}
\caption{$D$ meson and Kaon physics observables with the corresponding SM predictions and experimental bounds. Upper limits correspond to 95\% CL. \label{tab:obsK}}
\end{center} 
\end{table}

\begin{table}[p]
\begin{center}
\begin{tabular}{ | c | c | c | }
\hline
\textbf{Observable} & \textbf{SM prediction} & \textbf{Experimental bounds} \\
\hline
\hline
\cellcolor[gray]{0.92} $\Delta F=2$ processes& \cellcolor[gray]{0.92} &  \cellcolor[gray]{0.92} \cite{Gherardi:2020qhc} \\
 \hline
 $B^{0}-\overline{B}^{0}$: $|C_{B_d}^1|$ & 0 & $< 9.1\times 10^{-7}$ TeV$^{-2}$ \cite{Bona:2007vi,UTFIT:2016} \\
 \hline 
 $B_s^{0}-\overline{B}_s^{0}$: $|C_{B_s}^1|$ & 0 & $< 2.0\times 10^{-5}$ TeV$^{-2}$ \cite{Bona:2007vi,UTFIT:2016} \\
 \hline  
 $K^{0}-\overline{K}^{0}$: Re[$C_{K}^1$] & 0 & $< 8.0\times 10^{-7}$ TeV$^{-2}$ \cite{Bona:2007vi,UTFIT:2016} \\
 \hline
  $K^{0}-\overline{K}^{0}$: Im[$C_{K}^1$] & 0 & $< 3.0\times 10^{-9}$ TeV$^{-2}$ \cite{Bona:2007vi,UTFIT:2016} \\
 \hline    
 $D^{0}-\overline{D}^{0}$: Re[$C_{D}^1$] & 0 & $< 3.6\times 10^{-7}$ TeV$^{-2}$ \cite{Bona:2007vi,UTFIT:2016}  \\
 \hline
  $D^{0}-\overline{D}^{0}$: Im[$C_{D}^1$] & 0 & $< 2.2\times 10^{-8}$ TeV$^{-2}$ \cite{Bona:2007vi,UTFIT:2016} \\
 \hline    
 $D^{0}-\overline{D}^{0}$: Re[$C_{D}^4$] & 0 & $< 3.2\times 10^{-8}$ TeV$^{-2}$ \cite{Bona:2007vi,UTFIT:2016} \\
 \hline
  $D^{0}-\overline{D}^{0}$: Im[$C_{D}^4$] & 0 & $< 1.2\times 10^{-9}$ TeV$^{-2}$ \cite{Bona:2007vi,UTFIT:2016} \\
 \hline      
 $D^{0}-\overline{D}^{0}$: Re[$C_{D}^5$] & 0 & $< 2.7\times 10^{-7}$ TeV$^{-2}$ \cite{Bona:2007vi,UTFIT:2016} \\
 \hline
  $D^{0}-\overline{D}^{0}$: Im[$C_{D}^5$] & 0 & $< 1.1\times 10^{-8}$ TeV$^{-2}$ \cite{Bona:2007vi,UTFIT:2016} \\
\hline
\hline
\cellcolor[gray]{0.92} LFU in $\tau$ decays & \cellcolor[gray]{0.92} &\cellcolor[gray]{0.92} \cite{Gherardi:2020qhc} \\
\hline
$|g_\mu / g_e|^2$ & 1 & $1.0036\pm 0.0028$~\cite{Pich:2013lsa} \\
\hline
$|g_\tau / g_\mu |^2 $ & 1 & $1.0022\pm 0.0030$~\cite{Pich:2013lsa} \\
\hline
$|g_\tau / g_e|^2 $ & 1 & $1.0058\pm 0.0030$~\cite{Pich:2013lsa} \\
\hline
\hline
\cellcolor[gray]{0.92} LFV observables & \cellcolor[gray]{0.92} &  \cellcolor[gray]{0.92}\cite{Gherardi:2020qhc} \\
\hline
$\Br(\tau \to \mu \phi) $ & 0 & $ < 1.00\times10^{-7}$   \cite{Miyazaki:2011xe} \\ 
\hline   
$\Br(\tau \to 3\mu)$ & 0 &  $< 2.5\times 10^{-8} $     \cite{Hayasaka:2010np} \\     
\hline
$\Br(\tau\to \mu\gamma)$ & 0 & $ < 5.2\times10^{-8}$   \cite{Aubert:2009ag}\\  
\hline   
$\Br(\tau\to e\gamma)$ & 0 & $ < 3.9\times10^{-8}$   \cite{Aubert:2009ag} \\  
 \hline
 $\Br(\mu\to e\gamma)$ & 0 & $ < 5.0\times10^{-13}$   \cite{TheMEG:2016wtm} \\  
\hline   
$\Br(\mu \to 3 e)$ & 0 &  $< 1.2\times 10^{-12} $     \cite{Bellgardt:1987du} \\  
\hline
 $\Br_{\mu e}^{\text{(Ti)}}$ & 0 & $< 5.1 \times 10^{-12}$  ~\cite{Dohmen:1993mp} \\
 \hline
  $\Br_{\mu e}^{\text{(Au)}}$ & 0 & $< 8.3 \times 10^{-13}$   ~\cite{Bertl:2006up}  \\
\hline
\hline
 \cellcolor[gray]{0.92} EDMs & \cellcolor[gray]{0.92} & \cellcolor[gray]{0.92} \cite{Gherardi:2020qhc}  \\
\hline
$|d_e|$ & $< 10^{-44}\, \rm{e\cdot cm} $~\cite{Pospelov:2013sca,Smith:2017dtz} & $< 1.3 \times 10^{-29}\, \rm{e\cdot cm}$~\cite{Andreev:2018ayy}\\
\hline
$|d_\mu|$ & $< 10^{-42}\, \rm{e\cdot cm} $~\cite{Smith:2017dtz} & $< 1.9 \times 10^{-19}\, \rm{e\cdot cm}$~\cite{Bennett:2008dy} \\
\hline
$d_\tau$ & $< 10^{-41}\, \rm{e\cdot cm} $~\cite{Smith:2017dtz} & $(1.15\pm 1.70) \times 10^{-17}\, \rm{e\cdot cm}$\cite{Gherardi:2020qhc} \\
\hline
$d_n$ & $< 10^{-33}\, \rm{e\cdot cm} $~\cite{Khriplovich:1981ca}& $< 2.1 \times 10^{-26} \rm{e\cdot cm}$~\cite{Abel:2020gbr} \\
\hline
\hline 
\cellcolor[gray]{0.92} Anomalous & \cellcolor[gray]{0.92} &  \cellcolor[gray]{0.92}\cite{Gherardi:2020qhc} \\
 \cellcolor[gray]{0.92} Magnetic Moments & \cellcolor[gray]{0.92} &  \cellcolor[gray]{0.92} \\
\hline
$a_e-a_e^{SM}$ & $\pm 2.3 \times 10^{-13}$~\cite{Keshavarzi:2019abf,Parker:2018vye}& $(-8.9 \pm 3.6)\times 10^{-13}$~\cite{Hanneke:2008tm} \\
\hline
$a_\mu-a_\mu^{SM}$ & $\pm 43 \times 10^{-11}$ \cite{Aoyama:2020ynm} & $ (279 \pm 76)\times 10^{-11}$~\cite{Bennett:2006fi,Aoyama:2020ynm} \\
\hline
$a_\tau-a_\tau^{SM}$ & $\pm 3.9 \times 10^{-8}$~\cite{Keshavarzi:2019abf} & $(-2.1\pm 1.7)\times 10^{-7}$~\cite{Abdallah:2003xd} \\
\hline      
 \end{tabular}
\caption{Meson-mixing and leptonic observables and EDMs, with their SM predictions and experimental bounds. Upper limits correspond to 95\%CL. \label{tab:obs2}}
\end{center} 
\end{table}

\FloatBarrier

\begin{table}[t]
\begin{center}
\begin{tabular}{ | c | c | }
\hline
\textbf{Observable} & \textbf{Experimental bounds} \\
\hline
\hline
\cellcolor[gray]{0.92} $Z$ boson couplings & \cellcolor[gray]{0.92} App.~\ref{app:Zcouplings} \\\hline
	$\delta g^Z_{\mu_L}$ 	& $(0.3 \pm 1.1) 10^{-3}$	\cite{ALEPH:2005ab} \\ \hline
	$\delta g^Z_{\mu_R}$	& $(0.2 \pm 1.3) 10^{-3}$	\cite{ALEPH:2005ab} \\ \hline
	$\delta g^Z_{\tau_L}$	& $(-0.11 \pm 0.61) 10^{-3}$	\cite{ALEPH:2005ab} \\ \hline
	$\delta g^Z_{\tau_R}$	& $(0.66 \pm 0.65) 10^{-3}$	\cite{ALEPH:2005ab} \\ \hline
	$\delta g^Z_{b_L}$	& $(2.9 \pm 1.6) 10^{-3}$	\cite{ALEPH:2005ab} \\ \hline
	$\delta g^Z_{c_R}$	& $(-3.3 \pm 5.1) 10^{-3}$	\cite{ALEPH:2005ab} \\ \hline
	$N_\nu$	&	$2.9963 \pm 0.0074$	\cite{Janot:2019oyi} \\
\hline
\hline
\cellcolor[gray]{0.92} Drell-Yan  &  \cellcolor[gray]{0.92}  \\ \hline
	 $\sigma(pp \to \mu^+\mu^-)$	& \cite{Aaboud:2017buh,Angelescu:2018tyl} \\\hline
	 $\sigma(pp \to \tau^+\tau^-)$	& \cite{Aaboud:2017sjh,Angelescu:2018tyl} \\
\hline                       
\end{tabular}
\caption{ Limits on the deviations in $Z$ boson couplings to fermions from LEP I and from Drell-Yan high-energy tails at LHC. \label{tab:ZcouplLEP}}
\end{center} 
\end{table}

%%%%%%%%%%%%%%%%%%%%%%%%%%%%%%%%%
\appendix
%%%%%%%%%%%%%%%%%%%%%%%%%%%%%%%%%

%%%%%%%%%%%%%%%%%%%%%%%%%%%%%%%%%
\section{Analysis of observables and pseudo-observables}
\label{app:obs}
%%%%%%%%%%%%%%%%%%%%%%%%%%%%%%%%%

\subsection{$K_L\to \pi^0\nu\nu$ and $K^+\to \pi^+\nu\nu$}
\label{sec:Kpinunu}

The relevant parton level processes $s \to d \nu_\alpha \nu_\beta$ are described by the effective four-fermion Lagrangian
\be
	\LL_{\rm eff}^{d s\nu\nu} \supset \frac{4 G_F}{\sqrt{2}} \frac{\alpha}{4 \pi} V_{td}^* V_{ts} \sum_i C_i \OO_i~,
	\label{eq:LeffDF1bsnunu}
\ee 
with the following $\Delta F=1$ operators
\begin{align}
\OO_L^{ds\alpha\beta} &=\left(\bar{d} \gamma_\mu P_L s \right) ~ \left(\bar\nu_\alpha \gamma^\mu (1-\gamma_5) \nu_\beta \right) ~ ,
 \quad \OO_R^{ds\alpha\beta} = \left(\bar{d} \gamma_\mu P_R s \right) ~ \left(\bar\nu_\alpha \gamma^\mu (1-\gamma_5) \nu_\beta \right) ~ .
\label{eq:Opdsnunu}
\end{align}
The relations between the WCs of the operators above and the LEFT ones (in the basis of Ref.~\cite{Jenkins:2017jig}) are given by
\be
	[C_L]_{ds\alpha\beta} = N_{ds} \sum_{\alpha\beta}[L_{\nu d}^{V,LL}]_{\alpha\beta ds} ~ , \qquad
	[C_R]_{ds\alpha\beta} = N_{ds} \sum_{\alpha\beta}[L_{\nu d}^{V,LR}]_{\alpha\beta ds} ~ ,
\label{eq:Cbsnunu}
\ee
where $N_{ds} = (\sqrt{2} G_F \alpha V_{td}^* V_{ts} / \pi)^{-1}$.
Within the SM, the $K\to \pi \nu\nu$  decays are among the cleanest observables to determine the mixing between the top quark and the light generations as generated at loop level, as well as the magnitude of CP-violating (CPV) effects in the quark sector. The coefficients of the relevant four-fermion interaction are~\cite{Buras:1998raa,Buras:2015qea,Buras:2015yca}
\be
	[C_{L}]_{ds\alpha\beta}^{\SM} = -\frac{1}{s_W^2}\left(  \, X_t+ \frac{V_{cd}^* V_{cs}}{V_{td}^* V_{ts}} ~ X_c^\alpha \right)\delta_{\alpha\beta}~ , \qquad \qquad [C_{R}]_{ds\alpha\beta}^{\SM} =0~ ,
\ee
with $X_t=1.481$, $X_c^e=X_c^\mu=1.053\times 10^{-3}$, and $X_c^\tau=0.711\times 10^{-3}$. \par
In scenarios beyond the SM, these observables are highly sensitive to new sources of both CPC and CPV flavor mixing. In our case, at tree-level only $C_L$ is not vanishing. While at loop level a contribution to $C_R$ is generated (and included in the numerical analysis), it is suppressed by small down-quarks Yukawa couplings, and thus it is negligible.
The leading contributions to the WCs at $m_b$ scale, in terms of the UV parameters, are
\begin{align}
	[C_L]_{ds\alpha\beta}=&\, N_{ds}  \Big[ \frac{\lambda^{1L*}_{d \alpha} \lambda^{1L}_{s \beta}}{2 M_1^2} +  \frac{\lambda^{3L *}_{d \alpha} \lambda^{3L}_{s\beta}}{2 M_3^2}+\frac{1}{16 \pi^2}\frac{1}{12} \frac{m_t^2}{v^2} \Big[24 V^*_{t d} V_{ts} |V_{tb}|^2 \left(   \frac{\lambda^{3L *}_{b\alpha } \lambda^{3L}_{b\beta }}{M_3^2}\right)+ \nonumber\\
	& -3(3+2 \log (M^2/m_t^2)) \left(\left(  \frac{\lambda^{1L*}_{d\alpha } \lambda^{1L}_{k\beta }}{2 M_1^2} +  \frac{\lambda^{3L *}_{d\alpha } \lambda^{3L}_{k\beta }}{2 M_3^2}\right)V_{t s} V^*_{t k} + \right.  \nonumber \\
	&  \left. +\left(  \frac{\lambda^{1L*}_{k\alpha } \lambda^{1L}_{s\beta }}{2 M_1^2} +  \frac{\lambda^{3L *}_{k\alpha } \lambda^{3L}_{j\beta }}{2 M_3^2}\right)V_{t k} V^*_{t d}\right)\Big] \Big]+ \ldots~, \nonumber \\
	[C_R]_{ds\alpha \beta} \approx& ~0~.  \label{eq:EFTRnu}
\end{align}
In the $C_R\approx 0$ approximation, the branching ratios for the $K^+ \to \pi^+ \nu \nu$ and $K_L \to \pi^0 \nu \nu$ decays can be expressed in terms of the SM branching ratios, via a rescaling of the SM $C_L$ coefficient, in the following way
\begin{align}
&\Br(K^+ \to \pi^+ \nu \nu)= \Br(K^+ \to \pi^+ \nu_e \nu_e)_{\SM}\sum_{\alpha,\beta=1,2}\Big| \delta_{\alpha\beta}+\frac{[C_L]_{ds\alpha\beta}}{[C_{L}]_{ds11}^{\SM}}\Big|^2+ \\
&+\Br(K^+ \to \pi^+ \nu_\tau \nu_\tau)_{\SM} 
	\left[ \sum_{\alpha=1,2}\left(\Big|\frac{[C_L]_{ds\alpha 3}}{[C_{L}]_{ds33}^{\SM}}\Big|^2+\Big|\frac{[C_L]_{ds 3\alpha}}{[C_{L}]_{ds33}^{\SM}}\Big|^2\right) +\Big| 1+\frac{[C_L]_{ds33}}{[C_{L}]_{ds33}^{\SM}}\Big|^2 \right] , \nonumber\\
&\Br(K_L \to \pi^0 \nu \nu)=  \Br(K_L \to \pi^0 \nu \nu)_{\SM} \, \frac{1}{3} \,
	\Bigg[\sum_{\alpha,\beta=1,2}
		\left( \delta_{\alpha\beta}+\frac{\Im[N_{ds}^{-1} [C_L]_{ds\alpha\beta}]}{\Im[N_{ds}^{-1} [C_L]^{\SM}_{ds11}]} \right)^2+ \nonumber \\
&+\sum_{\alpha=1,2}\left(\left(\frac{\Im[N_{ds}^{-1} [C_L]_{ds\alpha 3}]}{\Im[N_{ds}^{-1} [C_{L}]_{ds33}^{\SM}]}\right)^2+\left(\frac{\Im[N_{ds}^{-1} [C_L]_{ds3\alpha}]}{\Im[N_{ds}^{-1} [C_{L}]_{ds33}^{\SM} ]}\right)^2\right) +\left( 1+\frac{\Im[N_{ds}^{-1} [C_L]_{ds33}]}{\Im[N_{ds}^{-1} [C_{L}]_{ds33}^{\SM}]}\right)^2 \Bigg]~,
\label{eq:BrKpinunu}
\end{align}
where \cite{Buras:2015qea,Buras:2015yca}
\be\begin{split}
	\Br(K^+ \to \pi^+ \nu_e \nu_e)_{\SM} &=3.06 \times 10^{-11}~, \\
	\Br(K^+ \to \pi^+ \nu_\tau \nu_\tau)_{\SM} &= 2.52 \times 10^{-11}~,\\
	\Br(K_L \to \pi^0 \nu \nu)_{\SM} &= 3.4 \times 10^{-11}~.
\end{split}\ee
The experimental bounds from NA62 \cite{NA62:2021zjw} and KOTO \cite{Tanabashi:2018oca} (see also \cite{KOTO:2020prk}) are
\be\begin{split}
	\Br(K^+ \to \pi^+ \nu\nu) & = (10.6^{+4.0}_{-3.5}) \times 10^{-11} ~,\\
	\Br(K_L \to \pi^0 \nu \nu) &<3.57 \times 10^{-9} \, \,  \, (95\% \text{CL}) ~.\\
\end{split}\ee
NA62 expects to reach a final sensitivity of about $10\%$ of the SM $K^+ \to \pi^+ \nu\nu$ breaching ratio \cite{Ruggiero:2017hjh}, while KOTO expects a future $95\%$CL upper limit for $\Br(K_L \to \pi^0 \nu \nu)$ of about $1.8 \times 10^{-10}$ at the end of stage-I. A proposed KOTO upgrade (stage-II) \cite{Aoki:2021cqa} or the KLEVER proposal at CERN \cite{Ambrosino:2019qvz} would be able to reach a final $20\%$ sensitivity of the SM rate.

Assuming that the contribution in $\nu_\tau$ is the dominant one we get the following approximate numerical expressions 
{\small \begin{align}
	10^{10} \Br(K^+ \to \pi^+ \nu \nu) &\approx 0.61+0.25 \left|1 
	+ (1.58 - i 0.51) \frac{\lambda^{1L *}_{d\tau} \lambda^{1L}_{s\tau}}{|V_{td}| |V_{ts}| m_1^2} + \right. \nonumber \\
	& \quad \left. + (1.45 - i 0.47) \frac{\lambda^{3L *}_{d\tau} \lambda^{3L}_{s\tau}}{|V_{td}| |V_{ts}| m_3^2} + \ldots \right|^2, \label{eq:Kpinunu} \\
	10^{10} \Br(K_L \to \pi^0 \nu \nu) &\approx 0.23 + 0.11 \left( 1 - \Im \left[5.4 \frac{\lambda^{1L *}_{d\tau} \lambda^{1L}_{s\tau}}{|V_{td}| |V_{ts}| m_1^2} + 4.9 \frac{\lambda^{3L *}_{d\tau} \lambda^{3L}_{s\tau}}{|V_{td}| |V_{ts}| m_3^2} \right] + \ldots  \right)^2 , \nonumber
\end{align}}
where $m_i=M_i/$TeV.

%--------------------------
\subsection{$K_{L(S)}\to \mu\mu$ and $K_L\to \pi^0 \ell\ell$}
\label{app:Kll}

The standard notation for the effective operators relevant to the $s\to d \ell^+ \ell^-$ rare processes is
\be
	\mathcal{H}_{\rm eff}^{sd\ell\ell} \supset - \frac{4 G_F}{\sqrt{2}} \frac{\alpha}{4 \pi s_W^2} \sum_i \mathcal{C}_i \OO_i~,
	\label{eq:LeffDF1}
\ee
where the $\mathcal{C}_i$ are evaluated at $m_s$ scale and the operators $\OO_i$ are defined as
\be\begin{array}{ll}
	\OO_9^{(\prime) sd\ell\ell } = (\bar s \gamma_\alpha P_{L (R)} d) (\bar \ell \gamma^\alpha \ell)~, \quad &
	\OO_{10}^{(\prime) sd\ell\ell } = (\bar s \gamma_\alpha P_{L (R)} d) (\bar \ell \gamma^\alpha \gamma_5 \ell)~, \\
	\OO_S^{(\prime) sd\ell\ell } =  (\bar s  P_{R (L)} d) (\bar \ell \ell)~, \quad &
	\OO_P^{(\prime) sd\ell\ell } =  (\bar s  P_{R (L)} d) (\bar \ell \gamma_5 \ell)~.
	\label{eq:Opdsll}
\end{array}\ee
The expressions for the WCs of the above operators in terms of the LEFT ones are 
\be\begin{array}{l l}
	\mathcal{C}_{9(10)}^{sd\ell\ell} = \frac{\NN_{sd}^{-1}}{2} \left( [L_{de}^{V,LR}]_{sd\ell\ell} \pm [L_{ed}^{V,LL}]_{\ell\ell sd} \right)~,  &
	\mathcal{C}_{9(10)}^{\prime sd\ell\ell} = \frac{\NN_{sd}^{-1}}{2} \left( [L_{ed}^{V,RR}]_{\ell\ell sd} \pm [L_{ed}^{V,LR}]_{\ell\ell sd} \right)~, \\
	\mathcal{C}_{S(P)}^{sd\ell\ell} = \frac{\NN_{sd}^{-1}}{2 } \left( [L_{ed}^{S,RR}]_{\ell\ell sd} \pm [L_{ed}^{S,RL}]^*_{\ell\ell sd} \right)~,  &
	\mathcal{C}_{S(P)}^{\prime sd\ell\ell} = \frac{\NN_{sd}^{-1}}{2 } \left( [L_{ed}^{S,RL}]_{\ell\ell sd} \pm [L_{ed}^{S,RR}]^*_{\ell\ell sd} \right)~.
	\label{eq:C910sdll}
\end{array}\ee
where $\NN_{sd} = \frac{4 G_F}{\sqrt{2}} \frac{\alpha}{4 \pi s_W^2} $.
In the Standard Model, the only non vanishing coefficients are \cite{Buchalla:1993wq}
\be
\mathcal{C}_{9,\rm SM}^{sd\ell\ell}= - \mathcal{C}_{10,\rm SM}^{sd\ell\ell}=  \left( V_{cs}^* V_{cd} Y_{NL}+ V_{ts}^* V_{td} Y(x_t) \right)~ .
\label{eq:C910sdSM}
\ee
with $Y_{NL}=3\times 10^{-4}$ and $Y(x_t)=0.94$.
In our analysis, the above coefficients are computed at one-loop level.
The short-distance contributions to the branching ratios for the leptonic $K_L\to \mu\mu$ and $K_S\to \mu\mu$ decays are given by
\be\begin{array}{l}
	\Br(K_L\to \mu^+\mu^-)_{SD} =
	\dfrac{\tau_{K_L} f_K^2  G_F^2 \alpha^2}{32\pi^3 m_K s_W^4}\sqrt{1-\dfrac{4m_\mu^2}{m_K^2}}  \\
	\quad \Big[ (m_K^2-4m_\mu^2) \left(\Re\left(\mathcal{C}_{S}^{ sd\mu\mu} - \mathcal{C}_{S}^{\prime sd\mu\mu}  \right)\dfrac{m_K^2}{m_s+m_d}\right)^2+ \\
	\quad + m_K^2\left(  \Re\left(\mathcal{C}_{10, \rm SM}^{ sd\mu\mu} +\mathcal{C}_{10}^{ sd\mu\mu}- \mathcal{C}_{10}^{\prime sd\mu\mu}  \right)2 m_\mu+ \Re\left(\mathcal{C}_{P}^{ sd\mu\mu} - \mathcal{C}_{P}^{\prime sd\mu\mu}  \right)\dfrac{m_K^2}{m_s+m_d}\right)^2\Big] ,\\ 
	 \Br(K_S\to \mu^+\mu^-)_{\rm SD} =
	 	\dfrac{\tau_{K_S} f_K^2  G_F^2 \alpha^2}{32\pi^3 m_K s_W^4}\sqrt{1-\dfrac{4m_\mu^2}{m_K^2}} \\
	 \quad \Big[ (m_K^2-4m_\mu^2) \left(\Im\left(\mathcal{C}_{S}^{ sd\mu\mu} - \mathcal{C}_{S}^{\prime sd\mu\mu}  \right)\dfrac{m_K^2}{m_s+m_d}\right)^2+ \\
	 \quad + m_K^2\left(  \Im\left(\mathcal{C}_{10, \rm SM}^{ sd\mu\mu} +\mathcal{C}_{10}^{ sd\mu\mu}- \mathcal{C}_{10}^{\prime sd\mu\mu}  \right)2 m_\mu+ \Im\left(\mathcal{C}_{P}^{ sd\mu\mu} - \mathcal{C}_{P}^{\prime sd\mu\mu}  \right)\dfrac{m_K^2}{m_s+m_d}\right)^2\Big] .
	\label{eq:BrKmumu}
\end{array}\ee
To these one should add the SM long-distance contribution, dominated by the intermediate $K_{L,S} \to \gamma^* \gamma^* \to \mu^+ \mu^-$ process. Calculating this in the SM proves to be very challenging (see e.g. Refs.~\cite{Isidori:2003ts,DAmbrosio:2017klp}) and while for the $K_S$ mode the long-distance component can be estimated to about $\approx 30\%$ accuracy, $\Br_{\rm LD} = \Br(K_S\to \mu^+\mu^-)_{\rm LD} = (5.18 \pm 1.50 )\times 10^{-12}$ \cite{DAmbrosio:2017klp}, for the $K_L$ mode the situation is worsened by the unknown sign of the amplitude $\cA(K_L \to \gamma\gamma)$.

For the rare semileptonic $K_L$ decays, we neglect the contributions from operators that are not generated at tree-level, which are also the ones not interfering with the SM interaction. Therefore, we apply a rescaling, with respect to SM, of the left-handed coefficient $\mathcal{C}_{9}^{sd\ell\ell}-\mathcal{C}_{10}^{sd\ell\ell}$
\be\begin{split}
    \Br(K_L\to \pi^0 \mu^+\mu^-)&=\Br(K_L\to \pi^0 \mu^+\mu^-)_{\text{SM}} \left(1 + \frac{\Im[\mathcal{C}_{9}^{sd\mu^+\mu^-}-\mathcal{C}_{10}^{sd\mu^+\mu^-}]}{\Im[\mathcal{C}_{9, \rm SM}^{sd\mu^+\mu^-}-\mathcal{C}_{10, \rm SM}^{sd\mu^+\mu^-}]}\right)^2~,\\
    \Br(K_L\to \pi^0 e^+e^-)&=\Br(K_L\to \pi^0 e^+e^-)_{\text{SM}} \left(1 + \frac{\Im[\mathcal{C}_{9}^{sde^+e^-}-\mathcal{C}_{10}^{sde^+e^-}]}{\Im[\mathcal{C}_{9, \rm SM}^{sde^+e^-}-\mathcal{C}_{10, \rm SM}^{sde^+e^-}]}\right)^2~,
\end{split}\ee
where the SM WCs are as in Eq.~\ref{eq:C910sdSM} and
\be\begin{split}
\Br(K_L\to \pi^0 \mu^+\mu^-)_{\text{SM}} &= (1.5\pm 0.3)\times 10^{-11} \quad \text{\cite{Isidori:2004rb} }~ , \\
\Br(K_L\to \pi^0 e^+e^-)_{\text{SM}} &= (3.2^{+ 1.2}_{-0.8})\times 10^{-11} \quad \text{\cite{Buchalla:2003sj} }~ .
\end{split}\ee
We evaluate these observables at one-loop level; however, in most of the regions of the parameter space, the tree-level contribution is dominant. At the tree-level, the branching fractions as a function of the model parameters are
\be\begin{split}
    \Br(K_L\to \mu^+\mu^-)_{SD} &\approx 8.4 \times 10^{-10}  \, \left( 1 - 1.50 \times 10^4 \, \Re\left[\frac{{\lambda^{3L}}^*_{s\mu}{\lambda^{3L}}_{d\mu}}{ M_3^2/\TeV^2}\right]\right)^2~ ,\\ 
	 \Br(K_S\to \mu^+\mu^-) &\approx  \Br_{\rm LD} + 1.86\times 10^{-13}  \, \left(1 + 4.23 \times 10^4 \,\Im\left[\frac{{\lambda^{3L}}^*_{s\mu}{\lambda^{3L}}_{d\mu}}{M_3^2/\TeV^2}\right]\right)^2 ~ , \\ 
    \Br(K_L\to \pi^0 \mu^+\mu^-) &\approx (1.5\pm 0.3)\times 10^{-11} \left(1 + 4.23 \times 10^4 \, \Im\left[\frac{{\lambda^{3L}}^*_{s\mu}{\lambda^{3L}}_{d\mu}}{ M_3^2/\TeV^2}\right]\right)^2 ~ , \\
    \Br(K_L\to \pi^0 e^+e^-) &\approx (3.2^{+ 1.2}_{-0.8})\times 10^{-11} \left(1 + 4.23 \times 10^4 \, \Im\left[ \frac{{\lambda^{3L}}^*_{se}{\lambda^{3L}}_{de}}{ M_3^2/\TeV^2} \right] \right)^2 ~ .	
\end{split}
\label{eq:BrKmumuTree}
\ee
The experimental bounds, at 95\% CL, on the studied Kaon branching fractions, whose expressions in the model are shown in the previous equations, are the following
\be\begin{split}
\Br(K_L\to \mu^+\mu^-)_{SD} & \lesssim 2.5 \times 10^{-9} \text{\cite{Isidori:2003ts}}~ , \\
\Br(K_S\to \mu^+\mu^-) & < 2.5 \times 10^{-10} \text{\cite{Aaij:2020sbt}} ~ , \\
\Br(K_L\to \pi^0\mu^+\mu^-) & < 4.5 \times 10^{-10}  \text{\cite{AlaviHarati:2000hs}}~ , \\
\Br(K_L\to \pi^0e^+e^-) & < 2.8 \times 10^{-10} \text{\cite{AlaviHarati:2003mr}}~ .
\end{split}\ee

%---------------------------------
\subsection{LFV in Kaon decays}
\label{app:KLmue}

The LFV Kaon decays are absent in SM; thus any non zero value for the $K_L\to \mu e$ branching ratio would be a signal of NP.
In terms of LEFT coefficients in the San Diego basis \cite{Jenkins:2017jig}, the branching ratios for the two leptonic channels $K_L\to \mu^-e+$ and $K_L\to e^-\mu^+$ decays are given by (see also Ref.~\cite{Angelescu:2020uug}):
{\small \begin{align}
	&\Br(K_L\to \mu^-e^+)=  \dfrac{\tau_{K_L} f_K^2 m_K m_\mu^2}{128\pi} \left(1-\dfrac{m_\mu^2}{m_K^2}\right)^2 \times \nonumber\\
	& \left[ \left| \left([L^{V,LR}_{ed}]^{\mu e s d} - [L^{V,LL}_{ed}]^{\mu e s d} + (s \leftrightarrow d) \right) - \dfrac{m_K^2}{m_\mu (m_s+m_d)} \left( [L^{S,RR}_{ed}]^{e \mu d s} - [L^{S,RL}_{ed}]^{e \mu d s} + (s \leftrightarrow d) \right)^* \right|^2 + \right. \nonumber\\
	&+ \left. \left| \left( [L^{V,RR}_{ed}]^{\mu e s d} - [L^{V,LR}_{de}]^{s d \mu e} + (s \leftrightarrow d) \right) - \dfrac{m_K^2}{m_\mu (m_s+m_d)} \left( [L^{S,RR}_{ed}]^{\mu e s d} - [L^{S,RL}_{ed}]^{\mu e s d} + (s \leftrightarrow d) \right) \right|^2  \right],\nonumber\\
    &\Br(K_L\to \mu^+e^-)=  \dfrac{\tau_{K_L} f_K^2 m_K m_\mu^2}{128\pi} \left(1-\dfrac{m_\mu^2}{m_K^2}\right)^2 \times \\
	& \left[ \left| \left([L^{V,LR}_{ed}]^{\mu e s d} - [L^{V,LL}_{ed}]^{\mu e s d} + (s \leftrightarrow d) \right) + \dfrac{m_K^2}{m_\mu (m_s+m_d)} \left( [L^{S,RR}_{ed}]^{\mu e s d} - [L^{S,RL}_{ed}]^{\mu e s d} + (s \leftrightarrow d) \right) \right|^2 + \right. \nonumber\\
	&+ \left. \left| \left( [L^{V,RR}_{ed}]^{\mu e s d} - [L^{V,LR}_{de}]^{s d \mu e} + (s \leftrightarrow d) \right) + \dfrac{m_K^2}{m_\mu (m_s+m_d)} \left( [L^{S,RR}_{ed}]^{e \mu d s} - [L^{S,RL}_{ed}]^{e \mu d s} + (s \leftrightarrow d) \right)^* \right|^2  \right],\nonumber
\end{align}}
where we approximated $m_\mu \gg m_e$.
In our analysis, the LEFT coefficients are computed at one-loop level. However, in the relevant parameter space the tree-level contribution from $S_3$ exchange is dominant. In this approximation we have
\be
	\Br(K_L\to \mu^-e^+) \approx \Br(K_L\to \mu^+e^-)  \approx  2.5 \times 10^{-2} ~ \left| \frac{ \lambda^{3L *}_{d\mu} \lambda^{3L}_{se} + \lambda^{3L *}_{s\mu} \lambda^{3L}_{de}}{m_3^2 } \right|^2~ ,
	\label{eq:BrKmueTree}
\ee
where $m_3=M_3/$TeV.
The experimental bound, at 95\% CL, on this LFV Kaon branching fraction is \cite{Ambrose:1998us} 
\be
    \Br(K_L\to \mu^\pm e^\mp)  < 4.7 \times 10^{-12} ~ .
\ee

The contributions to LFV Kaon decays $K^+ \to \pi^+ \mu^- e^+$ and $K^+ \to \pi^+ e^- \mu^+$ can be described following Refs.~\cite{Carrasco:2016kpy,Angelescu:2020uug}.
For simplicity, and since these will not be the most constraining observables in our model, we keep only the tree-level contribution from the $S_3$ exchange:
\be\begin{split}
    \Br(K^+\to \pi^+ \mu^- e^+) &\approx 1.1 \times 10^{-3} \left| \frac{{\lambda^{3L}_{d\mu}}^*\lambda^{3L}_{se}}{m_3^2 } \right|^2~, \\
     \Br(K^+\to \pi^+ e^- \mu^+) &\approx 1.1 \times 10^{-3} \left| \frac{{\lambda^{3L}_{de}}^*\lambda^{3L}_{s\mu}}{m_3^2 } \right|^2~. \\
\end{split}\ee
The recent NA62 constraint on the first decay is  $\Br(K^+\to \pi^+ \mu^- e^+) < 6.6 \times 10^{-11}$ at 90\% CL \cite{CortinaGil:2021mkg}. The best bound on the second decay mode is instead still from KTeV at BNL: $\Br(K^+\to \pi^+ e^- \mu^+) < 1.3 \times 10^{-11}$ at 90\% CL \cite{Abouzaid:2007aa}.

%%%%%%%%%%%%%%%%%%%%%%%%%%%%%%%%
\subsection{Rare $D$-meson decays}
\label{app:rareD}

While in general charm physics suffers from large uncertainties in the theory predictions, some channels offer good sensitivity to NP \cite{deBoer:2015boa,Fajfer:2015mia,Gisbert:2020vjx}. Particularly relevant to our setup are the $D^0 \to \ell^+ \ell^-$ and $D^+ \to \pi^+ \ell^+ \ell^-$ decays, which probe the FCNC $c \to u \ell^+ \ell^-$ transition.
These can be described by the following effective Hamiltonian at the $\mu_c = m_c$ scale:
\be
	\mathcal{H}_{\rm eff}^{c u \ell\ell} \supset - \frac{4 G_F}{\sqrt{2}} \frac{\alpha}{4 \pi} \left[ \sum_{i\neq T, T5} \left( \mathcal{C}_i \OO_i +  \mathcal{C}^\prime_i \OO^\prime_i \right) + \sum_{i = T, T5} C_i \OO_i ~ \right],
	\label{eq:HeffCharm}
\ee
with
\be\begin{array}{ll}
	\OO_9^{\ell (\prime)} = (\bar u \gamma_\alpha P_{L (R)} c) (\bar \ell \gamma^\alpha \ell)~, \quad &
	\OO_{10}^{\ell (\prime)} = (\bar u \gamma_\alpha P_{L (R)} c) (\bar \ell \gamma^\alpha \gamma_5 \ell)~, \\[1pt]
	\OO_S^{\ell (\prime) } = (\bar u  P_{R (L)} c) (\bar \ell \ell)~, \quad &
	\OO_P^{\ell (\prime) } = (\bar u  P_{R (L)} c) (\bar \ell \gamma_5 \ell)~. \\[1pt]
	\OO_7^{ (\prime) } = \frac{m_c}{e} (\bar u  \sigma_{\mu\nu} P_{R (L)} c) F^{\mu\nu}~, \quad &
	\OO_8^{(\prime) } = \frac{m_c g_s}{e^2} (\bar u  \sigma_{\mu\nu} T^a P_{R (L)} c) G_a^{\mu\nu}~, \\[1pt]
	\OO_{T (T5)}^\ell = \frac{1}{2} (\bar u  \sigma_{\mu\nu} c) (\bar \ell \sigma^{\mu\nu} (\gamma_5) \ell)~. &
	\label{eq:OpCharm}
\end{array}\ee
In the SM the only short-distance contributions arise along the $\mathcal{C}_7$ and $\mathcal{C}^\ell_9$ coefficients (see e.g. \cite{Gisbert:2020vjx} and references therein), while long-distance contributions are very challenging to evaluate due to $m_c \sim \Lambda_{\rm QCD}$ and can reliably be computed only on the lattice.
The purely leptonic decays are dominated by  the long-distance contribution $D^0 \to \gamma^* \gamma^* \to \ell^+ \ell^-$, which gives a branching ratio estimated to be of order $\sim 10^{-11}$. On the other hand the present experimental constraints at 95\%CL are still larger by several orders of magnitude:
\be
	\Br(D^0 \to e^+ e^-) < 7.9 \times 10^{-8}~, ~~
	\Br(D^0 \to \mu^+ \mu^-) < 6.2 \times 10^{-9}~, ~~
	\Br(D^0 \to \mu^\pm e^\mp) < 1.3 \times 10^{-8}~.
\ee
For this reason, when setting limits on NP one can thus safely neglect the SM contribution, writing the branching ratio as:
\be\begin{split}
	\Br(D^0 \to \mu^+ \mu^-) =& \frac{G_F^2 \alpha^2 m_{D^0}^5 f_{D^0}^2}{64 \pi^3 m_c^2 \Gamma_D} \sqrt{1 - \frac{4 m_\mu^2}{m_{D^0}^2}} \Bigg[ \left(1 - \frac{4 m_\mu^2}{m_D^2} \right) \left| \mathcal{C}_S^\mu - \mathcal{C}_S^{\mu \prime} \right|^2 +  \\
	& +   \left| \mathcal{C}_P^{\mu} - \mathcal{C}_P^{\mu \prime} + \frac{2 m_\mu m_c}{m_{D^0}^2} \left( \mathcal{C}_{10}^{\mu} - \mathcal{C}_{10}^{\mu \prime} \right) \right|^2 \Bigg]~,
\end{split}\ee
where $f_{D^0} \approx 206.7$MeV, $m_{D^0} \approx 1864.83$MeV, $\tau_{D^0} \approx 4.101 \times 10^{-13}$s.

Also in the semileptonic decay $D^+ \to \pi^+ \ell^+ \ell^-$ one can neglect the SM contribution, given the present experimental accuracy. The 95\%CL limit on the muonic channel gives the constraint \cite{deBoer:2015boa,Gisbert:2020vjx}:
\be\begin{split}
 &1.3 |\mathcal{C}_7^+ |^2 + 1.3 |\mathcal{C}_9^{\mu +}|^2 + |\mathcal{C}_{10}^{\mu +}|^2 + 2.6 |\mathcal{C}_S^{\mu +}|^2 + 2.7 |\mathcal{C}_P^{\mu +}|^2 + 0.4 |\mathcal{C}_T^{\mu}|^2 + 0.4 |\mathcal{C}_{T5}^{\mu}|^2 + \\
 &+ 0.3 \Re \Big[ \mathcal{C}_9^{\mu +}  \mathcal{C}_T^{\mu *} \Big] + 1.1 \Re \Big[ \mathcal{C}_{10}^{\mu +}  \mathcal{C}_P^{\mu + *} \Big] + 2.6 \Re \Big[ \mathcal{C}_7^{+}  \mathcal{C}_9^{\mu + *} \Big] + 0.6 \Re \Big[ \mathcal{C}_7^+  \mathcal{C}_T^{\mu *} \Big] < 1~,
\end{split}\ee
where $\mathcal{C}_i^+ = \mathcal{C}_i + \mathcal{C}^\prime_i$.
In terms of the LEFT coefficients at the $m_c$ scale, the $\mathcal{C}_i$'s are given by:
\be\begin{array}{ll}
	\mathcal{C}_{9}^{\ell} = \frac{1}{2 \mathcal{N}_{cu}} \left( [L^{V,LR}_{ue}]_{uc\ell\ell} +  [L^{V,LL}_{eu}]_{\ell\ell uc} \right)~, \quad &
	\mathcal{C}_{10}^{\ell} = \frac{1}{2 \mathcal{N}_{cu}} \left( [L^{V,LR}_{ue}]_{uc\ell\ell} -  [L^{V,LL}_{eu}]_{\ell\ell uc} \right)~, \\[5pt]
	\mathcal{C}_{9}^{\ell \prime} = \frac{1}{2 \mathcal{N}_{cu}} \left( [L^{V,RR}_{eu}]_{\ell\ell uc} +  [L^{V,LR}_{eu}]_{\ell\ell uc} \right)~, \quad &
	\mathcal{C}_{10}^{\ell \prime} = \frac{1}{2 \mathcal{N}_{cu}} \left( [L^{V,RR}_{eu}]_{\ell\ell uc} -  [L^{V,LR}_{eu}]_{\ell\ell uc} \right)~, \\[5pt]
	\mathcal{C}_{7} = \frac{1}{\mathcal{N}_{cu}}\frac{e}{m_c} [L_{u\gamma}]_{uc}~,  \quad &
	\mathcal{C}_{7}^\prime = \frac{1}{\mathcal{N}_{cu}} \frac{e}{m_c} [L_{u\gamma}]_{cu}^*~, \\[5pt]
	\mathcal{C}_{S} = \mathcal{C}_{P} = \frac{1}{2 \mathcal{N}_{cu}} [L^{S,RR}_{eu}]_{\ell\ell uc}~,  \quad &
	\mathcal{C}_{S}^\prime = - \mathcal{C}_{P}^\prime = \frac{1}{2 \mathcal{N}_{cu}} [L^{S,RR}_{eu}]_{\ell\ell cu}^*~, \\[5pt]
	\mathcal{C}_{T}^{\ell \prime} = \frac{1}{\mathcal{N}_{cu}} \left( [L^{T,RR}_{eu}]_{\ell\ell uc} +  [L^{T,RR}_{eu}]_{\ell\ell cu}^* \right)~, \quad &
	\mathcal{C}_{T5}^{\ell \prime} = \frac{1}{\mathcal{N}_{cu}} \left( [L^{T,RR}_{eu}]_{\ell\ell uc} -  [L^{T,RR}_{eu}]_{\ell\ell cu}^* \right)~,
\end{array}\ee
where $\mathcal{N}_{cu} = \frac{4 G_F}{\sqrt{2}} \frac{\alpha}{4 \pi}$.
While in our numerical analysis we keep all the one-loop contributions to these coefficients arising from the two LQs (including one-loop matching and RGE), for illustrative purposes we report here only the tree-level contributions, which are expected to give the dominant terms. In this limit, the only non-vanishing coefficients are:
 \be\begin{split}
 	 [L^{V,LL}_{eu}]_{\ell\ell uc}^{(0)} &=  V_{u i} V_{c j}^* \left( \frac{\lambda^{1L *}_{i \ell} \lambda^{1L}_{j \ell}}{2 M_1^2} + \frac{\lambda^{3L *}_{i \ell} \lambda^{3L}_{j \ell}}{2 M_3^2} \right)~, \\
	 [L^{V,RR}_{eu}]_{\ell\ell uc}^{(0)} &=  \frac{\lambda^{1R *}_{u \ell} \lambda^{1R}_{c \ell}}{2 M_1^2} ~, \\
	 [L^{S,RR}_{eu}]_{\ell\ell uc}^{(0)} &= - 4  [L^{T,RR}_{eu}]_{\ell\ell uc}^{(0)} =  - V_{u i} \frac{\lambda^{1L *}_{i \ell} \lambda^{1R}_{c \ell}}{2 M_1^2} ~.
\end{split}\ee

%%%%%%%%%%%%%%%%%%%%%%%%%%%%%%%%%
\subsection{$b\to d \mu \mu$ and $b\to d ee$ decays}
\label{app:bdmumu}

We consider the leptonic branching ratios
\be
	\Br(B^0 \to \mu ^+ \mu ^-)~, \quad \quad	\Br(B^0 \to e ^+ e ^-)~,
\ee
and the semileptonic ones
\be
	\Br(B^+ \to \pi^+  \mu ^+ \mu ^-)~, \quad \quad	\Br(B^+ \to \pi^+  e^+ e^-) ~ .
\ee
At partonic level, these processes are induced by the same low-energy operators as $b \to d \mu \mu$ and $b \to d e e$, respectively. Among these operators, the ones that in our framework can arise at tree-level are
\be\begin{array}{ll}
	\OO_9^{ db\ell\ell } = (\bar d \gamma_\alpha P_{L (R)} b) (\bar \ell \gamma^\alpha \ell)~, \quad &
	\OO_{10}^{ db\ell\ell } = (\bar d \gamma_\alpha P_{L (R)} b) (\bar \ell \gamma^\alpha \gamma_5 \ell)~.
	\label{eq:Opbdmumu}
\end{array}\ee
where $\ell = \mu,e$. The expressions for the WCs of the above operators in terms of the LEFT ones are 
\be\begin{array}{l l}
	\mathcal{C}_{9(10)}^{db\ell\ell} = \frac{\NN_{db}^{-1}}{2} \left( [L_{de}^{V,LR}]_{db\ell\ell} \pm [L_{ed}^{V,LL}]_{\ell\ell db} \right), 
	\label{eq:C910bdmumu}
\end{array}\ee
where $\NN_{db} = \frac{4 G_F}{\sqrt{2}} \frac{\alpha}{4 \pi} V_{td}^* V_{tb}$.\\ 

In our model, the tree-level generated effective vertices lead to left-handed four-fermion interactions with $\mathcal{C}^9_{d b \ell\ell} = - \mathcal{C}^{10}_{d b \ell\ell}$, which are the ones that interfere with the SM processes. The effective Lagrangian for these processes also contains scalar ($\OO_S^{(\prime) db\ell\ell }$) and pseudoscalar ($\OO_P^{(\prime) db\ell\ell }$) semileptonic operators, as well as dipoles ($\OO_7^{(\prime) db}$) and other two vector four-fermion operators ($\OO_{9(10)}^{\prime db\ell\ell }$). However, their contributions to the $b\to d \mu\mu (ee)$ transitions is negligible in the $S1+S3$ model with respect to the one coming from the operators in Eq.~\eqref{eq:Opbdmumu}. Therefore, we take into account only the $\mathcal{C}^9_{d b \ell\ell}$ and $\mathcal{C}^{10}_{d b \ell\ell}$ contributions to the considered $B$-mesons decays, including however, in the numerical analysis, also the one-loop corrections to these coefficients.
The branching ratios are given then by,
\be\begin{split}
\Br(B^0 \to \ell^- \ell^+)&=\Br(B_{d}\to \ell^- \ell^+)_{\text{SM}} \left|1 + \frac{\mathcal{C}^{10}_{d b \ell \ell}}{{\mathcal{C}^{10\,\text{SM}}_{d b \ell \ell}}}\right|^2~,\\
\Br(B^+\to \pi^+ \ell ^+ \ell ^-) &=\Br(B^+\to \pi^+ \ell ^+ \ell ^-)_{\text{SM}}\left|1 + \frac{\mathcal{C}^{9}_{d b \ell \ell}-\mathcal{C}^{10}_{d b \ell \ell}}{{\mathcal{C}^{9\,\text{SM}}_{d b \ell \ell}}-{\mathcal{C}^{10\,\text{SM}}_{d b \ell \ell}}}\right|^2~,
\end{split}\ee
where~\cite{Khodjamirian:2017fxg} 
\be
\mathcal{C}^{9,SM}_{d b \ell \ell} = - \mathcal{C}^{10,SM}_{d b \ell \ell} = 4.2 ~.
\ee
The SM predictions are~\cite{Bobeth:2013uxa, Bailey:2015nbd},
\be\begin{split}
	\Br(B^0 \to \mu \mu)_{\text{SM}}& =  (1.06 \pm 0.09) \times 10^{-10}~,  \\
	\Br(B^+\to \pi^+ \mu^- \mu^+)_{\text{SM}}& = (2.04\pm 0.21) \times 10^{-8}~, \\
	\Br(B^0\to ee)_{\text{SM}}& = (2.48\pm 0.21) \times 10^{-15}~, \\
	\Br(B^+\to \pi^+ e^- e^+)_{\text{SM}}& = (2.04\pm 0.24) \times 10^{-8}~.
\end{split}\ee
The branching ratios are measured for the case of the muons~\cite{Chatrchyan:2013bka,Aaij:2015nea},
\be\begin{split}
	\Br(B^0 \to \mu \mu)_{\text{exp}}& =  (1.1 \pm 1.4) \times 10^{-10}~,  \\
	\Br(B^+\to \pi^+ \mu^- \mu^+)_{\text{exp}}& = (1.83\pm 0.24) \times 10^{-8}~,
\end{split}\ee
while for the electrons only upper bounds exists so far~\cite{Tanabashi:2018oca},
\be\begin{split}
	\Br(B^0 \to ee)_{\text{exp}}& <  8.3 \times 10^{-8}~,  \\
	\Br(B^+\to \pi^+ e^- e^+)_{\text{exp}}& < 8 \times 10^{-8}~. 
\end{split}\ee
Numerically, at tree-level, we get:
\be
\dfrac{\Br(B^0\to \ell^- \ell^+)}{\Br(B^0\to \ell^- \ell^+)_{\text{SM}}}
    =\dfrac{\Br(B^+\to \pi^+\ell^- \ell^+)}{\Br(B^+\to \pi^+\ell^- \ell^+)_{\text{SM}}}
    \approx \left| 1 + 139.23  \frac{\lambda^{3L }_{b \ell} \lambda^{3L *}_{d \ell}}{ M_3^2/\TeV^2} \right|^2~.
\ee

%%%%%%%%%%%%%%%%%%%%%%%%%%%%%%%%%
\subsection{$\mu$--$e$ conversion in nuclei}
\label{app:mu2e}

Searches for $\mu$--$e$ conversion in nuclei yield competitive LFV constraints on LQ interactions coupling light quarks to electrons and muons. The relevant observable is the branching ratio
\begin{equation}
	\rm BR_{\mu e} = \frac{\Gamma_{\rm{conversion}}}{\Gamma_{\rm{capture}}}~,
\end{equation}
where $\Gamma_{\rm{conversion}}$ denotes the rate of conversion of a muon captured by the $1s$ nucleus $A$ to an electron $\mu A \to e+A$ and $\Gamma_{\rm{capture}}$ the normalising rate $\mu A \to \nu_\mu + A$. \par
The current upper bounds are 
\begin{equation}
	\rm BR^{(\text{Ti})}_{\mu e} < 4.3 \times 10^{-12} \quad \quad \rm{and} \quad \quad \rm BR^{(\text{Au})}_{\mu e} < 7 \times 10^{-13} \quad (90\% \quad \rm{CL})~,
\end{equation}
as set by experiments on titanium~\cite{Dohmen:1993mp} and gold targets~~\cite{Bertl:2006up}, respectively. 
The COMET (JPARC) \cite{Kuno:2013mha} and Mu2e (Fermilab) \cite{Ankenbrandt:2006zu,Knoepfel:2013ouy,Bartoszek:2014mya} experiments are expected to reach the $10^{-16}$ sensitivity, while the PRISM proposal at JPARC aims at the $10^{-18}$ level.

Numerical calculations of the relevant transition matrix elements in the different nuclei~\cite{Kitano:2002mt} yield the following formula for the conversion rates
\be\begin{split}
  \label{eq:conversion-rate}
  \Gamma_{\text{conversion}} &= 2 G_F^2 
  \left| A_R^* D +  (2g_{LV}^{(u)} + g_{LV}^{(d)}) V^{(p)} + (g_{LV}^{(u)} + 2 g_{LV}^{(d)}) V^{(n)}  \right. \\
  &\qquad\qquad\left.+  (G_S^{(u,p)} g_{LS}^{(u)} + G_S^{(d,p)} g_{LS}^{(d)}+ G_S^{(s,p)} g_{LS}^{(s)}) S^{(p)}  \right. \\
  &\qquad\qquad\left. + (G_S^{(u,n)} g_{LS}^{(u)} + G_S^{(d,n)} g_{LS}^{(d)} + G_S^{(s,n)} g_{LS}^{(s)}) S^{(n)} \right|^2 \\
  &\quad+ (L\leftrightarrow R) ~.
\end{split}\ee
The numerical factors that are the same for both nuclei are $G_S^{(u,p)} = G_S^{(d,n)} = 5.1$, $G_S^{(d,p)} = G_S^{(u,n)} = 4.3$, and $G_S^{(s,p)} = G_S^{(s,n)} = 2.5$~\cite{Kosmas:2001mv} and the ones that are different are presented in Table~\ref{tbl:TabTiAuData}. The various expressions containing WCs are given in the LEFT basis by
\be\begin{split}
  g_{LV}^{(q)} = \frac{\sqrt{2}}{G_F} \left( [L_{eq}^{V,LL}]_{e\mu qq} + [L_{qe}^{V,LR}]_{qq e\mu} \right)~,& \quad
	 g_{RV}^{(q)} = \frac{\sqrt{2}}{G_F} \left( [L_{eq}^{V,RR}]_{e\mu qq} + [L_{eq}^{V,LR}]_{e\mu qq} \right)~, \\
	g_{LS}^{(q)} = \frac{\sqrt{2}}{G_F} \left( [L_{eq}^{S,RR}]_{e\mu qq} + [L_{eq}^{S,RL}]_{e\mu qq} \right)~,& \quad
	g_{RS}^{(q)} = \frac{\sqrt{2}}{G_F} \left( [L_{eq}^{S,RR}]_{\mu e qq}^* + [L_{eq}^{S,RL}]_{\mu e qq}^* \right)~, \\
	A_L = \frac{\sqrt{2} e}{4 G_F m_\mu} [L_{e\gamma}]_{\mu e}~,& \quad
	A_R = \frac{\sqrt{2} e}{4 G_F m_\mu} [L_{e\gamma}]_{\mu e}^*~.
\end{split}\ee
In the model these receive both tree-level and loop contributions and, since the tree-level ones are typically suppressed by small couplings to 1st generation, loop effects are important and the expressions are quite cumbersome. Expanding numerically these coefficients, the terms with largest numerical pre-factor are ($m_i \equiv M_i / \TeV$):
\be\begin{array}{ll}
    g_{LV}^u \approx 0.065 \frac{V_{ui}^* V_{uj} \lambda^{1L *}_{i e}\lambda^{1L}_{i \mu}}{m_1^2} + 0.060 \frac{V_{ui}^* V_{uj} \lambda^{3L *}_{i e}\lambda^{3L}_{i \mu}}{m_3^2} + \ldots~, &
    g_{LV}^{d(s)} \approx  0.12 \frac{\lambda^{3L *}_{d(s) e}\lambda^{3L}_{d(s) \mu}}{m_3^2} + \ldots ~, \\
    g_{RV}^u \approx 0.062 \frac{\lambda^{1R *}_{u e}\lambda^{1R}_{u \mu}}{m_1^2} + \ldots~, & 
    g_{RV}^{d(s)} \approx 0~, \\
    g_{LS}^u \approx - 0.11 \frac{V_{ui}^* \lambda^{1L *}_{i e}\lambda^{1R}_{u \mu}}{m_1^2} + \ldots~, & 
    g_{LS}^{d(s)} \approx 0~, \\
    g_{RS}^u \approx - 0.11 \frac{V_{ui} \lambda^{1L}_{i \mu}\lambda^{1R *}_{u e}}{m_1^2} + \ldots~, & 
    g_{RS}^{d(s)} \approx 0~, \\
    A_L \approx 0.055 \frac{\lambda^{1L *}_{b \mu}\lambda^{1R}_{t e}}{m_1^2} (1 + \log m_1^2) + \ldots~, & \\
    A_R \approx 0.055 \frac{\lambda^{1L}_{b e}\lambda^{1R *}_{t \mu}}{m_1^2} (1 + \log m_1^2) + \ldots~. & \\
\end{array}\ee

\begin{table}
{\small
  \begin{center}
  \begin{tabular}{lccc} \hline
    Nucleus                        &$D [10^{-9}\rm GeV]$  & $V^{(p)} [10^{-9}\rm GeV]$ & $V^{(n)} [10^{-9}\rm GeV] $\\ \hline
    $\text{\text{Ti}}^{48}_{22}$    & 9.91 &              4.54    &             5.37  \\
    $\text{\text{Au}}^{197}_{79}$ &  21.68 &              11.17
                                                                                 &             16.75    \\ 
&&&\\
  & $S^{(p)} [10^{-9}\rm GeV]$ & $S^{(n)} [10^{-9}\rm GeV] $ &
                                                         $\Gamma_{\text{capture}}[10^6 s^{-1}]$
    \\ \hline
 $\text{\text{Ti}}^{48}_{22}$ &   4.22 & 4.99            &  2.59\\
 $\text{\text{Au}}^{197}_{79}$& 7.04 &  10.53        &  13.07\\
\hline
\end{tabular}
  \end{center}
}%END \small
  \caption{Data relevant for the $\mu \to e $ conversion in nuclei ~\cite{Kitano:2002mt}.}
  \label{tbl:TabTiAuData}
\end{table}

%%%%%%%%%%%%%%%%%%%%%%%%%%%%%%%%%
\subsection{$Z$ boson couplings}
\label{app:Zcouplings}

The couplings of the $Z$ boson have been measured very precisely at LEP 1. At one-loop, the LQ model generates contributions to these very well measured quantities, which pose strong constraints on the model.
The RGE-induced contributions in models aimed at addressing the $B$-anomalies have first been studied in Ref.~\cite{Feruglio:2016gvd,Feruglio:2017rjo,Cornella:2018tfd}. Here we include the effect of finite corrections from the one-loop matching.

The pseudo-observables corresponding to the effective couplings of the $Z$ boson to fermions, $\mathcal{G}^{Z}_{\psi}$, are defined from the residues of the $Z$ pole in the $e^+ e^- \to \psi \bar{\psi}$ scattering amplitude (see e.g. \cite{ALEPH:2005ab} and references therein).
In general, both in the SM and in the SMEFT, radiative corrections induce absorptive parts to these pseudo-observables, which are therefore complex parameters. In the standard approach, used also by LEP to set limits on BSM contributions to the pole couplings, absorptive parts from the SM are included in the SM prediction, while it is assumed that BSM contributions are real, by defining real effective couplings as $g^Z_\psi = \Re(\mathcal{G}^{Z}_{\psi})$ \cite{ALEPH:2005ab}.\footnote{When working at leading-order, the $Z$-pole coupings pseudo-observables can be put in correspondence to these effective Lagrangian couplings: $\mathcal{L}_{eff} = - \frac{g}{c_\theta} [g^{Z}_{\psi}]_{ij} Z_\mu (\bar \psi_i \gamma^\mu \psi_j)$.}
This is justified by the fact that absorptive parts due to BSM corrections are typically much smaller than the corresponding real parts and below the experimental sensitivity (we also show this explicitly numerically below).

One can extract the SM contribution to the effective couplings as \mbox{$g^Z_\psi = [g^{Z}_{\psi}]^{\SM} + \delta g^Z_\psi$}, where in the SM at tree-level one has $[g^{Z}_{\psi}]_{ij}^{\SM, \text{tree}} = \delta_{ij} (T_{3L}^\psi - Q_\psi s_\theta^2)$. The measurements of these pseudo-observables and the predictions for the SM contributions can be found in \cite{ALEPH:2005ab}.
Also often used is the effective number of neutrino species at the $Z$ peak, which depends on the couplings to neutrinos as
\be
	N_\nu = \sum_{\alpha\beta} \left| \delta_{\alpha\beta} + \frac{[\delta g^Z_{\nu}]_{\alpha\beta}}{ [g^{Z}_{\nu}]^{\SM}} \right|^2~,
\ee
where $[g^{Z}_{\nu}]^{\SM} \approx 0.502$. The latest update on the extraction of $N_\nu$ from LEP data is given in \cite{Janot:2019oyi}. We collect in Table~\ref{tab:ZcouplLEP} the limits used in our fit.

Working at one-loop accuracy, there are two possible contributions to these pseudo-observables in our setup: tree-level contributions from the SMEFT operators (which include one-loop matching from the UV model at a scale $M$ and RGE from $M$ to the scale $\mu$)\footnote{There might be some indirect contributions via modifications to $G_F$, but are negligible in our model.} and one-loop matrix elements for $Z \to \psi \bar{\psi}$ from the tree-level generated semileptonic operators. The result is:
\be\begin{split}
	[\delta g^Z_{e_L}]_{\alpha\beta}  \approx&   - v^2 \Re\left[ \frac{1}{2} [C^{(1+3)}_{H\ell}]^{(1)}_{\alpha\beta}(\mu) + \frac{N_c}{16\pi^2} [C_{\ell q}^{(1 - 3)}]^{(0)}_{\alpha\beta i i} I_L^{\overline{MS}}(u_i,m_Z^2, \mu) \right] ~,\\
	[\delta g^Z_{e_R}]_{\alpha\beta}  \approx& - v^2 \Re\left[ \frac{1}{2} [C_{He}]^{(1)}_{\alpha\beta}(\mu) + \frac{N_c}{16\pi^2} [C_{eu}]^{(0)}_{\alpha\beta i i} I_R^{\overline{MS}}(u_i,m_Z^2, \mu) \right] ~, \\
	[\delta g^Z_{\nu}]_{\alpha\beta}  \approx& - v^2 \Re\left[ \frac{1}{2} [C^{(1-3)}_{H\ell}]^{(1)}_{\alpha\beta}(\mu) + \frac{N_c}{16\pi^2} [C_{\ell q}^{(1 + 3)}]^{(0)}_{\alpha\beta i i} I_L^{\overline{MS}}(u_i,m_Z^2, \mu) \right] ~, \\
	[\delta g^Z_{d_L}]_{i j}  \approx&   - \frac{v^2}{2}  \Re\left[ [C^{(1+3)}_{Hq}]^{(1)}_{i j}(m_Z) \right] ~,\\
	[\delta g^Z_{u_L}]_{i j}  \approx&   - \frac{v^2}{2}  \Re\left[ V_{i k} V^*_{j l} [C^{(1-3)}_{Hq}]^{(1)}_{k l}(m_Z)) \right] ~,\\
	[\delta g^Z_{u_R}]_{i j}  \approx&   - \frac{v^2}{2}  \Re\left[ [C_{Hu}]^{(1)}_{i j}(m_Z) \right] ~,\\
	[\delta g^Z_{d_R}]_{i j}  \approx&   - \frac{v^2}{2}  \Re\left[ [C_{Hd}]^{(1)}_{i j}(m_Z) \right] ~,
\end{split}\ee
where $\mu$ is the scale at which the operators are evaluated, $C^{(1\pm 3)}_X = C^{(1)}_X \pm C^{(3)}_X$, and we included only one-loop matrix elements from up-type quark loops, since the top quark gives the dominant effect and we want to also describe the effect of possible large couplings to $c_R$.
The expressions for the one-loop matching of the SMEFT operators to the LQ model can be found in \cite{Gherardi:2020det}. The loop functions are given by
\be\begin{split}
	 v^2 I_L^{\overline{MS}}(u_i, q^2, \mu) &= \frac{1}{9} \Big[ 5 q^2 (1 - 3 s_\theta^2) - 6 m_{u_i}^2 (1+6 s_\theta^2) + \\
		& \qquad - 3 \left(m_{u_i}^2 - q^2 + 3 (2 m_{u_i}^2 + q^2) s_\theta^2 \right) \text{DiscB}(q^2, m_{u_i}, m_{u_i}) + \\
		& \qquad + 3 (-3 m_{u_i}^2 + q^2 - 3 q^2 s_\theta^2) \log \frac{\mu^2}{m_{u_i}^2} \Big]~, \\
	 v^2 I_R^{\overline{MS}}(u_i, q^2, \mu) &= -\frac{5}{3} q^2 s_\theta^2 + m_{u_i}^2 (2 - 4 s_\theta^2) + \\
		& \qquad + \left(m_{u_i}^2 - (2 m_{u_i}^2 + q^2) s_\theta^2 \right) \text{DiscB}(q^2, m_{u_i}, m_{u_i}) + \\
		& \qquad + ( m_{u_i}^2 - q^2 s_\theta^2) \log \frac{\mu^2}{m_{u_i}^2} \Big]~, \\
	\text{DiscB}(q^2, m, m) &= \frac{\sqrt{q^2(q^2 - 4 m^2)}}{q^2} \log \frac{2 m^2 -q^2 + \sqrt{q^2(q^2 - 4 m^2)}}{2 m^2}~.
\end{split}\ee
As expected, the dependence on the arbitrary scale $\mu$ cancels between the logarithm inside the $I_{L/R}$ integrals and the RGE terms present inside the one-loop coefficients $[C]^{(1)}(\mu)$. We note that in case of light quarks, the logarithm inside the DiscB function combines with the $\log \mu^2/m_{u_i}^2$ terms to give $\approx \log \mu^2 / m_Z^2$, as expected. 
Numerically, one has $I_L^{\overline{MS}}(c ,m_Z^2, \mu = 1.5 \TeV) = 0.102 + i 0.044$, $I_R^{\overline{MS}}(c ,m_Z^2, \mu = 1.5 \TeV) = - 0.23 - i 0.10$ while $I_L^{\overline{MS}}(t ,m_Z^2, \mu = 1.5 \TeV) = - 1.91$ and $I_R^{\overline{MS}}(t ,m_Z^2, \mu = 1.5 \TeV) = 1.83$. Furthermore, in the limit $q^2 \to 0$,
\be
	I_{R,L}^{\overline{MS}}( t, 0, \mu) \approx \pm \frac{m_t^2}{v^2} \log \frac{\mu^2}{m_Z^2} = \pm 1.95  \log \left( \frac{\mu^2}{1.5 \TeV^2} \right)~.
\ee
From this is clear that the contributions from top loops (purely real) are a factor of $\sim 10$ larger than contributions from charm (that generates also absorptive terms).

When all contributions are included, and normalizing LQ masses to 1 TeV by defining $m_i \equiv M_i / \TeV$, numerical expressions are:
\begin{align}
	10^3 \delta g^Z_{e_{\alpha L}}  \approx&  0.80 \frac{(\lambda^{3L}_{b\alpha})^2}{m_3^2} (1 + 0.35 \log m_3^2) +0.59 \frac{(\lambda^{1L}_{b\alpha})^2}{m_1^2}(1 + 0.39 \log m_1^2) + \nonumber \\
	& + 0.11 \frac{(\lambda^{3L}_{s\alpha})^2}{m_3^2}- 0.10 \frac{(\lambda^{1L}_{s\alpha})^2}{m_1^2}+\ldots , \\
	10^3 \delta g^Z_{\nu_{\alpha}}  \approx&  1.31 \frac{(\lambda^{3L}_{b\alpha})^2}{m_3^2} (1 + 0.37 \log m_3^2) + 0.11 \frac{(\lambda^{1L}_{b\alpha})^2}{m_1^2}   + 0.11 \frac{(\lambda^{1L}_{s\alpha})^2}{m_1^2}-0.16 \frac{\lambda^{3L}_{b\alpha} \lambda^{3L}_{s\alpha}}{m_3^2} + \ldots , \\
	10^3 \delta g^Z_{e_{\alpha R}}  \approx&  -0.67 \frac{(\lambda^{1R}_{t\alpha})^2}{m_1^2} (1 + 0.37 \log m_1^2) + 0.059 \frac{(\lambda^{1R}_{c\alpha})^2}{m_1^2} + 0.030 \frac{(\lambda^{1R}_{u\alpha})^2}{m_1^2}  + \ldots , \\
	10^3 \delta g^Z_{b_{L}}  \approx&  - 0.044 \frac{(\lambda^{1L}_{b \tau})^2 + (\lambda^{1L}_{b \mu})^2}{m_1^2} + \ldots , \\
	10^3 \delta g^Z_{c_{R}}  \approx&  - 0.014 \frac{(\lambda^{1R}_{c \tau})^2 + (\lambda^{1R}_{c \mu})^2}{m_1^2} + \ldots , 
\end{align}
where the dots represent smaller, thus negligible, contributions given the present experimental sensitivity.

\subsection{$\epsilon_K'/\epsilon_K$}
\label{app:epsprime}

The observable $\epsilon_K'/\epsilon_K$ is strongly suppressed within the SM, not only due to the CKM hierarchy but also to 
	an accidental SM low-energy property (the so-called $\Delta I=1/2$ rule). Consequently this observable is a particularly sensitive probe 
	of non-standard sources of both CP and flavor symmetry breaking. The experimental world average~\cite{Tanabashi:2018oca} reads 
\begin{equation}
\left(\epsilon_K'/\epsilon_K\right)_{\rm exp} = (16.6 \pm 2.3) \times 10^{-4} ~ .
\end{equation}
The main difficulty in making predictions for $\epsilon_K'/\epsilon_K$ resides with the theory side. The uncertainty on the SM calculation is plagued by long-distance effects on the hadronic matrix elements of the relevant operators. The value we use in the present work is taken from~\cite{Gisbert:2018tuf}
\begin{equation}
\left(\epsilon_K'/\epsilon_K\right)_{\rm SM} \approx (15 \pm 7) \times 10^{-4},
\end{equation}	
and it is also in good overall agreement with other recent results~\cite{Cirigliano:2019obu,Aebischer:2019mtr}. \par
The master formula~\cite{Aebischer:2020jto} for the NP contribution is given by
\begin{equation}\label{eq:epsprime_NP}
\left(\epsilon_K'/\epsilon_K\right)_{\rm NP} = \sum_i P_i(\mu_{\rm ew}) \rm Im \left[ C_i(\mu_{\rm ew}) - C'_i(\mu_{\rm ew}) \right].
\end{equation}
In Table \ref{tab:P_i} we list all the relevant operators $Q_i$ that are non-vanishing in our framework, together with their WCs $C_i$ and the respective $P_i$ values. The relations between these WCs and the LEFT ones are given by
\begin{align}
&C_{VLL}^{u^i}= [L_{ud}^{V1,LL}]_{u^iu^isd}~, \qquad C_{VLR}^{u^i}= [L_{du}^{V1,LR}]_{sdu^iu^i}~, \qquad \tilde{C}_{VLL}^{u^i}= \frac{1}{2} [L_{ud}^{V8,LL}]_{u^iu^isd}~,  \notag \\
&\tilde{C}_{VLR}^{u^i}= \frac{1}{2} [L_{du}^{V8,LR}]_{sdu^iu^i}~, \qquad C_{VLL}^{d^i}= [L_{dd}^{V,LL}]_{sdd^id^i}~, \qquad
C_{VLR}^{d^i}= [L_{dd}^{V1,LR}]_{sdd^id^i}~, \notag \\
&C_{SLR}^{d^i}= - [L_{dd}^{V8,LR}]_{d^idsd^i}~. 
\end{align}	
where $u^i$ can be the $u$ or $c$ quark and $d^i$ the $d$, $s$ or $b$ quark. The $C'_i$ are the WCs of the corresponding chirality-flipped operators obtained by interchanging $P_L \leftrightarrow P_R$. Keeping only the terms with the largest numerical pre-factors in our model, we find the approximate expression ($m_i \equiv M_i / \TeV$):
\begin{align}
\left(\epsilon_K'/\epsilon_K\right)_{\rm NP} &\approx  10^{-6}\frac{2.4((\lambda_{t\mu}^{1R})^2+(\lambda_{t\tau}^{1R})^2)-0.3(\lambda_{c\mu}^{1R} \lambda_{t\mu}^{1R}+\lambda_{c\tau}^{1R} \lambda_{t\tau}^{1R})}{m_1^2} \notag \\
&+10^{-8}\frac{ ((\lambda_{s\mu}^{1L})^2+(\lambda_{s\tau}^{1L})^2-2\lambda_{s\tau}^{1L}\lambda_{u\tau}^{1L})\log m_1^2}{m_1^2} + 10^{-8}\frac{1.1((\lambda_{s\mu}^{3L})^2+(\lambda_{s\tau}^{3L})^2)(1+\log m_3^2)}{m_3^2}~.
\end{align}	

\begin{table}[t]
\centering
\begin{tabular}{|lc|}
\hline
	\qquad \qquad \qquad $Q_i$
				 & 
	$P_i$
\\
\hline\hline
  
$Q_{VLL}^u = (\bar s^i \gamma_\mu P_L d^i)(\bar u^j \gamma^\mu P_L u^j)$                & $-3.3  \pm 0.8$  \\
$Q_{VLR}^u = (\bar s^i \gamma_\mu P_L d^i)(\bar u^j \gamma^\mu P_R u^j)$                & $-124  \pm 11$   \\
$\tilde{Q}_{VLL}^u = (\bar s^i \gamma_\mu P_L d^j)(\bar u^j \gamma^\mu P_L u^i)$           & $1.1   \pm 1.2$   \\
$\tilde{Q}_{VLR}^u = (\bar s^i \gamma_\mu P_L d^j)(\bar u^j \gamma^\mu P_R u^i)$           & $-430  \pm 40$  \\[0.1cm]
$Q_{VLL}^d = (\bar s^i \gamma_\mu P_L d^i)(\bar d^j \gamma^\mu P_L d^j)$                & $1.8   \pm 0.5$  \\
$Q_{VLR}^d = (\bar s^i \gamma_\mu P_L d^i)(\bar d^j \gamma^\mu P_R d^j)$                & $117   \pm 11$   \\
$Q_{SLR}^d = (\bar s^i P_L d^i)(\bar d^j P_R d^j)$                                      & $204   \pm 20$   \\[0.1cm]
$Q_{VLL}^s = (\bar s^i \gamma_\mu P_L d^i)(\bar s^j \gamma^\mu P_L s^j)$                & $0.1   \pm 0.1$  \\
$Q_{VLR}^s = (\bar s^i \gamma_\mu P_L d^i)(\bar s^j \gamma^\mu P_R s^j)$                & $-0.17 \pm 0.04$ \\
$Q_{SLR}^s = (\bar s^i P_L d^i)(\bar s^j P_R s^j)$                                      & $-0.4  \pm 0.1$  \\[0.1cm]
$Q_{VLL}^c = (\bar s^i \gamma_\mu P_L d^i)(\bar c^j \gamma^\mu P_L c^j)$                & $0.5   \pm 0.1$  \\
$Q_{VLR}^c = (\bar s^i \gamma_\mu P_L d^i)(\bar c^j \gamma^\mu P_R c^j)$                & $0.8   \pm 0.1$  \\
$\tilde{Q}_{VLL}^c = (\bar s^i \gamma_\mu P_L d^j)(\bar c^j \gamma^\mu P_L c^i)$           & $0.7   \pm 0.1$  \\
$\tilde{Q}_{VLR}^c = (\bar s^i \gamma_\mu P_L d^j)(\bar c^j \gamma^\mu P_R c^i)$           & $1.3   \pm 0.2$ \\[0.1cm]
$Q_{VLL}^b = (\bar s^i \gamma_\mu P_L d^i)(\bar b^j \gamma^\mu P_L b^j)$                & $-0.33 \pm 0.03$ \\
$Q_{VLR}^b = (\bar s^i \gamma_\mu P_L d^i)(\bar b^j \gamma^\mu P_R b^j)$                & $-0.22 \pm 0.03$ \\
$\tilde{Q}_{VLL}^b = (\bar s^i \gamma_\mu P_L d^j)(\bar b^j \gamma^\mu P_L b^i)$           & $0.3   \pm 0.1$  \\
$\tilde{Q}_{VLR}^b = (\bar s^i \gamma_\mu P_L d^j)(\bar b^j \gamma^\mu P_R b^i)$           & $0.4   \pm 0.1$  \\
\hline
\end{tabular}
\caption{
The operators corresponding to the WCs entering the master formula for NP effects in Eq. \eqref{eq:epsprime_NP} (and are non-vanishing in the $S_1 + S_3$ framework) and the respective $P_i$ values. 
}
  \label{tab:P_i}
\end{table}

%%%%%%%%%%%%%%%%%%%%%%%%%%%%%%%%
\subsection{Neutron EDM}
\label{nEDM}

Electric dipole moments (EDMs) can be strong probes of CP-violating effects in leptoquark scenarios \cite{Dekens:2018bci,Altmannshofer:2020ywf}. We discussed lepton EDMs constraints for $S_1$ and $S_3$ in \cite{Gherardi:2020qhc} and focus here on the neutron EDM.
EDMs are CP-odd observables whose measurements can set bounds on the imaginary parts of the model parameters. Neutron EDM, in particular, receives contributions from the short range QCD interactions involving the EDMs $d_q$ and chromo-EDMs (cEDMs) $\hat{d}_q$ of the light quarks ($u$, $d$ and $s$) that are valence quarks of the nucleon. These (c)EDMs are defined by the following Lagrangian
\be
 \mathcal{L}^{(c)EDM}=-i\frac{d_q}{2} \, \bar{q}\sigma^{\mu\nu}\gamma_5 q \, F_{\mu\nu}-i\frac{\hat{d}_q}{2} \, \bar{q}\sigma^{\mu\nu}T^a\gamma_5 q \, G_{\mu\nu}^a ~ .
\ee
The quark dipole moments can be expressed in terms of the LEFT WCs
\be\begin{split}
d_u=-\Im\left(2 [L_{u\gamma}]_{uu}(1{\rm GeV})\right)~ , & \quad \hat{d}_u=-\Im\left(2 [L_{uG}]_{uu}(1{\rm GeV})\right)~ , \\
d_d=-\Im\left(2 [L_{d\gamma}]_{dd}(1{\rm GeV})\right)~ , & \quad \hat{d}_d=-\Im\left(2 [L_{dG}]_{dd}(1{\rm GeV})\right)~ ,\\
d_s=-\Im\left(2 [L_{d\gamma}]_{ss}(1{\rm GeV})\right)~ ,& \quad \hat{d}_s=-\Im\left(2 [L_{dG}]_{ss}(1{\rm GeV})\right)~ ,
\end{split}\ee
where 1GeV is the neutron mass scale. One should notice, however, that in our model there is not any imaginary part for the $L_{d\gamma(dG)}$ coefficients, as a consequence of the fact that the $S_1$ LQ is not coupled to right-handed down quarks. 
In principle, the neutron EDM receives a contribution from the CP-odd gluon Weinberg operator ($\OO_{3\widetilde{G}}$) as well. However, in our model this operator arises at two-loops level.

The neutron EDM can be expressed in terms of the quark (c)EMDs as follows
\be\begin{split}
d_n = &-\frac{v}{\sqrt{2}}\left( \beta_n^{uG} \,  \hat{d}_u + \beta_n^{dG}\,  \hat{d}_d  +\beta_n^{u\gamma} \, d_u +\beta_n^{d\gamma} \, d_d +\beta_n^{s\gamma} \, d_s\right)= \\
&= -\frac{v}{\sqrt{2}}\left( \beta_n^{uG} \,  \hat{d}_u  +\beta_n^{u\gamma} \, d_u \right)~,
\end{split}\ee
where the $\beta_n^{qV}$ are the hadronic matrix elements and in the second line we have taken into account the fact that in our model the down quark dipoles are vanishing. For the cEDMs the $\beta_n^{qG}$ factors are estimated using QCD sum rules and they are $ \beta_n^{uG}=-2^{+1.5}_{-0.8}\times 10^{-4} e \, {\rm fm}$ and $\beta_n^{dG} = -4^{+3.1}_{-1.7}\times 10^{-4} e \, {\rm fm}$ \cite{Hisano:2012sc,Engel:2013lsa}; one can notice that there are currently affected by a large uncertainty, bigger than $50\%$. The contributions from the strange quark cEDM are cancelled if a Peccei-Quinn mechanism is used to remove the contribution from the QCD $\theta$-term \cite{Pospelov:2000bw}. The $\beta_n^{q\gamma}$ matrix elements are evaluated with lattice QCD \cite{Bhattacharya:2015esa,Bhattacharya:2015wna,Bhattacharya:2016zcn,Gupta:2018lvp} and their most recent estimates are $-\frac{v}{\sqrt{2}} \beta_n^{u\gamma}=-0.204^{+0.011}_{-0.010}$, $-\frac{v}{\sqrt{2}} \beta_n^{d\gamma}=+0.784\pm 0.028$ and $-\frac{v}{\sqrt{2}} \beta_n^{s\gamma}=-0.0027\pm 0.0016$.

The experimental bound, on the absolute value of the neutron EDM, at $90\%$ CL is~\cite{Abel:2020gbr}
\be
|d_n|<1.8 \times 10^{-26} {\rm e \cdot cm}~.
\ee
In our analysis, for simplicity we take into account the central values for the hadronic matrix elements. 
We find that EDMs give negligible contributions in constraining the model once the other observables entering the fit are taken into account.

%%%%%%%%%%%%%%%%%%%%%%%%%%%%%%%%
%%%%%%%%%%%%%%%%%%%%%%%%%%%%%%%%
{\small
\bibliography{Biblio}
\bibliographystyle{JHEP}}

\end{document}